\documentclass[aps,prd,10pt,showpacs,amsmath,showkeys,twocolumn,floatfix,amssymb, preprintnumbers, nofootinbib, superscriptaddress]{revtex4-1} 
\usepackage{epsfig,dcolumn}
\usepackage{graphicx}
\usepackage{comment} 
\usepackage[usenames]{color}
\usepackage{graphicx}
\usepackage{bm}
 \usepackage{ifpdf}

\usepackage[normalem]{ulem}
\usepackage[dvipsnames]{xcolor}
\usepackage[utf8]{inputenc}
\usepackage{hyperref}
\hypersetup{ 
  pdfnewwindow=true,      
  colorlinks=true,        
  linkcolor=PineGreen,    
  citecolor=PineGreen,    
  filecolor=PineGreen,    
  urlcolor=PineGreen      
}

\newcommand{\be}{\begin{eqnarray}}
\newcommand{\ee}{\end{eqnarray}}

\newcommand{\cor}[2]{{\color{red} \sout{#1}} {\color{blue} #2}}

\begin{document}

\title{ A lattice model of heavy-light three-body system}

\author{Peng~Guo}
\email{pguo@csub.edu}

\affiliation{Department of Physics and Engineering,  California State University, Bakersfield, CA 93311, USA}
\affiliation{Kavli Institute for Theoretical Physics, University of California, Santa Barbara, CA 93106, USA}

\author{Michael~D\"oring}
\email{doring@gwu.edu}
\affiliation{Department of Physics, The George Washington University, Washington, DC 20052, USA}
\affiliation{Theory Center, Thomas Jefferson National Accelerator Facility, Newport News, VA 23606, USA}

\date{\today}

\begin{abstract} 
We present a study of a $1+1$ dimensional heavy-light three-body system in finite volume. The heavy-light system is simulated by a coupled-channel $\phi^4$ type lattice model, and   both ground state and excited states of multiparticle energy spectra   are measured on various lattices. The lattice simulation data analysis is performed based on variational approach.
  \end{abstract} 


\maketitle

\section{Introduction}
\label{intro}

Much of strong interaction phenomenology manifests itself in few-body systems. Due to the many degrees of freedom their quantitative description is more complicated than in the two-body case. Lattice QCD calculations, e.g., of multi-pion systems~\cite{Beane:2007es, Detmold:2008fn, Horz:2019rrn}, provide an ab-initio understanding of few-body dynamics. However, the dynamics is encoded in a discrete spectrum of energy eigenvalues corresponding to the cubic volume with usually periodic boundary conditions in which these calculations are performed. 
In the  past few years,  much progress  toward mapping such spectra to infinite-volume amplitudes has been made \cite{Kreuzer:2008bi,Kreuzer:2009jp,Kreuzer:2012sr,Polejaeva:2012ut,Briceno:2012rv,Hansen:2014eka,Hansen:2015zga,Hansen:2016fzj,Hammer:2017uqm,Hammer:2017kms,Meissner:2014dea,Briceno:2017tce,Sharpe:2017jej,Mai:2017bge,Doring:2018xxx, Romero-Lopez:2018rcb,Guo:2016fgl,Guo:2017ism,Guo:2017crd,Guo:2018ibd,Guo:2018xbv,Guo:2019hih,Blanton:2019igq, Romero-Lopez:2019qrt}.   
Some approaches provide connections to the infinite volume without fully resolving the few-body dynamics~\cite{Agadjanov:2016mao, Bulava:2019kbi, Hansen:2017mnd, Hashimoto:2017wqo}.
In first applications of infinite-volume mappings, the two- and three-body lattice QCD spectra by the NPLQCD collaboration~\cite{Beane:2007es, Detmold:2008fn} was analyzed in Ref.~\cite{Mai:2018djl}, and the recent data of Ref.~\cite{Horz:2019rrn} was analyzed/predicted in Ref.~\cite{Blanton:2019vdk} and Ref.~\cite{Mai:2019fba}, respectively. 

Most of these developments are along the line of building connections between reaction amplitudes in infinite volume and long-range correlations due to  the periodic structure of a finite box.
To a certain extent, these developments  may be regarded as extensions of  the  L\"uscher formula \cite{Luscher:1990ux,Rummukainen:1995vs,Christ:2005gi,He:2005ey,Lage:2009zv,Doring:2011vk,Doring:2011ip,Aoki:2011gt,Briceno:2012yi,Hansen:2012tf,Guo:2012hv,Guo:2013vsa}. In the two-particle sector, L\"uscher's formula  demonstrates a clear separation of two physical scales: (i) short-range physics in a single box that is  parametrized by  a scattering amplitude, and  (ii) long-distance correlations in a periodic lattice structure that is described by zeta functions ~\cite{Luscher:1990ux}.  Although formulating quantization conditions by using   reaction amplitudes as input to producing discrete energy spectra or vice versa   presents a more conventional foundation of dealing with multiparticle  dynamics, one is confronted to deal simultaneously with questions regarding  infinite and finite-volume physics. As pointed out in Ref.~\cite{Guo:2019hih},  the quantization condition of multiparticle system in finite volume may be constructed directly from Faddeev equation-type coupled-channel integral equations. The discretized   finite-volume wave functions may be treated as coefficients of secular equations, hence, the quantization condition given by the determinant of secular equations is free of infinite volume reaction amplitudes, and is presented in terms of finite-volume Green's function and particle interactions.  By avoiding the most difficult part of multiparticle   dynamics, the quantization condition   may be more efficient for practical data analysis of lattice  calculation  results. 

The aim of this work is to demonstrate the feasibility of the  variational approach, and explore  a robust form of a quantization condition which depends  only on the lattice structure and interaction potentials.  The interaction potentials may be parameterized and treated as  inputs to fit the lattice simulation results. Once the parameters of  interaction potentials  are extracted,  in principle, all the physics information is  complete, and the dynamical information, such as scattering amplitudes, may be computed in a separate step. 

In the present work, using a non-relativistic heavy-light three-body system  with short-range pair-wise interactions as a pedagogical example, we show in details that the three-body problem in finite volume may be turned into a coupled two-body like homogeneous Lippmann-Schwinger equations. The quantization conditions may be obtained in terms of only three-body finite-volume free Green's function and interaction potentials, and are free of the specific form of finite-volume variational basis functions.  In the second part of this work, a coupled-channel $\phi^4$ theory lattice model is used to simulate the heavy-light three-body system proposed in this work. The multiple levels of multiparticle spectra are extracted from lattice simulation. In order to compensate finite lattice spacing, finite volume and relativistic dynamics in lattice simulation,  the non-relativistic formalism derived in the first part of the presentation is thus reformulated to a relativistic lattice version for the practical fit of lattice simulation results. At last, the quantization conditions are applied to extract the parameters of the heavy-light three-body system: the mass of particles and the coupling strength of short-range interactions.

The paper is organized as follows. Using a heavy-light system as a specific example,  the   variational approach is explained in detail  in Section \ref{3bdynamics}. A lattice model of the heavy-light system,  using a Monte Carlo updating algorithm  of  the   lattice model,  and the construction of multi-particle operators and multi-particle spectra in lattice simulation    are described in Section \ref{phi4sols}.    The strategy of   data analysis, reformulating the finite-volume heavy-light three-body system to a relativistic lattice version and numerical results   are presented  in Section  \ref{data}. The summary and outlook are given in Section \ref{summary}.

\section{A non-relativistic  heavy-light three-body system in finite volume}\label{3bdynamics}

Although the non-relativistic formalism  may not be the most suitable framework for describing  relativistic lattice simulations,  it still provides a clean and simple presentation of physics in a finite box, and is able to capture all the key features of finite-volume dynamics   without extra complication of relativistic effect. Therefore, in this Section, all the presentations are given in  a non-relativistic framework.

 We start our discussion with the simple example of a two-body finite volume system in $3D$. The  wave function of two identical bosons in the center-of-mass frame satisfies the homogeneous Lippmann-Schwinger equation,
\begin{equation}
\Phi (\mathbf{ r}; q) = \int_{L^3} d \mathbf{ r}' G_{0}^{L}(\mathbf{ r}- \mathbf{ r}'; q ) V(r') \Phi (\mathbf{ r}'; q), \label{2bLuscher}
\end{equation}
where $q$ stands for the relative momentum of the two identical particles and periodic the Green's function satisfies 
\begin{equation}
\left ( q^2 + \nabla^2 \right ) G_{0}^{L}(\mathbf{ r} ; q ) = \sum_{\mathbf{ n} \in \mathbb{Z}^3} \delta ( \mathbf{ r} + \mathbf{ n} L).
\end{equation}
The  wave function in   finite volume has to be periodic as well, such as, $\Phi (\mathbf{ r} + \mathbf{ n} L; q)  = \Phi (\mathbf{ r}; q) $. 
According to the variational approach \cite{Guo:2019hih}, the solution of the homogeneous Eq.~(\ref{2bLuscher}) is given by the linear superposition of independent solutions in infinite volume, $\Phi_q (\mathbf{ r}) = \sum_{[J]} c_{[J]} \Psi_{[J]} (r; q) Y_{[J]}  (\mathbf{ \hat{r}}) $, where $\Psi_{[J]}$ stands for  the infinite-volume partial-wave scattering solution of the Schr\"odinger equation.  
After partial wave projection  Eq.(\ref{2bLuscher}) yields
\begin{align}
&  \sum_{[J]} c_{[J]}  \Psi_{[J]} (r; q)   \nonumber \\
&  =\sum_{[J']} c_{[J']}   \int_{0}^R {r'}^2 d r'    G_{[J],[J']}^{L}( r,r'; q )   V(r') \Psi_{[J']} (r'; q), \label{2bLuscherpwa}
\end{align}
where
\begin{equation}
 G_{[J],[J']}^{L}( r,r'; q )   = \int d \mathbf{\hat{ r}} d \mathbf{\hat{ r}}'  Y^*_{[J]}  (\mathbf{ \hat{r}}) G_{0}^{L}(\mathbf{ r}- \mathbf{ r}'; q )   Y_{[J']}  (\mathbf{ \hat{r}}').
 \end{equation}
Using the asymptotic form of the infinite-volume wave function,
\begin{equation}
\Psi_{[J]} (r; q) \sim (4\pi) i^J  \left [ j_J(qr) + i t_J(q) h_J^{(+)} (q r) \right ],
\end{equation}
where $  t_J (q) = - \frac{q}{(4\pi)i^J}   \int_{0}^R r^2 d r  j_J(q r) V(r) \Psi_{[J]} (r; q) $ is the partial-wave scattering amplitude,
and   also performing the   partial wave expansion    of $G_{[J],[J']}^L$,
\begin{align}
&  G_{[J],[J']}^{L}( r,r'; q )  \nonumber \\
& \stackrel{r>r'}{  =} q  \left [  \delta_{[J], [J']}  n_J (qr) -  \mathcal{M}_{[J'], [J]} (q) j_J (qr) \right ] j_{J'} (q r'),
\end{align}
the homogeneous  Eq.~(\ref{2bLuscherpwa}) leads to the  L\"uscher formula \cite{Luscher:1990ux}, $\det \mathcal{D} (q) =0$, where  
 \begin{equation}
  \mathcal{D}_{[J], [J']}  (q)=  \delta_{[J], [J']}  \left ( i + \frac{1}{t_J (q)}  \right )-  \mathcal{M}_{[J'], [J]} (q) .
 \end{equation}
 The equivalent derivation in differential form of the variational approach by constructing a finite-volume wave function in terms of an infinite-volume wave function  can be found in Ref.~\cite{Guo:2018ibd}. Using the infinite-volume wave function solutions as   basis functions  shows some advantages in the two-body sector: (i) the quantization condition is given in terms of physical on-shell scattering amplitudes; (ii) the relation between finite-volume solutions and infinite-volume solutions is clearly demonstrated.

 However,  in case of multiparticle interaction, imposing physical constraints, such as  the asymptotic form  of the infinite-volume wave functions, and unitarity of multiparticle scattering amplitudes, means an additional difficulty on top of the task of obtaining the quantization condition itself. All the  difficulties in multiparticle interaction in finite volume in fact  starts with an ambitious goal  at the very beginning: expressing quantization conditions in terms of infinite scattering amplitudes or wave functions, and handling both infinite-volume and finite-volume dynamics simultaneously.  The multiparticle interaction in the infinite volume alone is already difficult to solve.

 In this work, we explore the possibility of separating finite-volume   and infinite volume dynamics; the two scenarios are connected by interaction potentials instead of scattering amplitudes. In this way, the quantization condition may be expressed in terms of periodic lattice structure and potentials alone, and is free of multiparticle dynamics in the infinite volume. The multiparticle dynamics in infinite volume may be computed separately  once the information of the potentials  is extracted from  fits to lattice results. The proposed approach is far less ambitious, however, it provides a way of avoiding dealing with scattering amplitudes, and a much more practical formalism for just the purpose of lattice data fitting.

Before we move on to the three-body problem, we would like to demonstrate the key idea of our proposal by using the finite-volume  two-body problem as a simple example.  Starting the from homogeneous equation, Eq.(\ref{2bLuscher}), again,  according to variational approach \cite{Guo:2018ibd,Guo:2019hih},  the continuous equation, Eq.(\ref{2bLuscher}), may be turned into a homogenous matrix equation if the wave function is expanded in terms of certain basis functions, and then   the quantization condition may be obtained by the determinant condition of this matrix  equation. In principle, the discrete energy spectra do not depend on the specific choice of basis functions. In Ref.~\cite{Guo:2018ibd}, the basis functions were constructed explicitly, but  the specific form of finite-volume wave functions is not the focus of the present work.   Alternatively,  the quantization condition may be obtained from the homogeneous Lippmann-Schwinger equation directly, Eq.~(\ref{2bLuscher}) or Eq.~(\ref{2bLuscherpwa}). One way of numerically solving the LS equation is by discretizing it in coordinate space, in a single box,
 \begin{equation}
\sum_j \left [ \delta_{ i, j } - w_j  G_{0}^{L}(\mathbf{ r}_i- \mathbf{ r}_j; q ) V(r_j) \right ] \Phi_j =0,   
\end{equation}
 where $\mathbf{ r}_i$ and  $w_j$ denote the chosen coordinates and associated integration weights in a single box.  Hence, the non-trivial solutions exist,  provided that the determinant of the homogeneous equation vanishes, $\det D (q) =0$, where
 \begin{equation}
D_{i,j} (q) = \delta_{ i, j } - w_j  G_{0}^{L}(\mathbf{ r}_i- \mathbf{ r}_j; q ) V(r_j)   \ .  \label{2bquantcorr}
\end{equation}
The determinant condition,  Eq.~(\ref{2bquantcorr}), is equivalent to the L\"uscher formula and yields the discrete energy spectra.

The quantization condition,  Eq.~(\ref{2bquantcorr}),  may also be transferred into  momentum space representation by  using the  Fourier transformation relation in finite volume,
\begin{equation}
 \widetilde{\Phi }(\mathbf{ p}; q)  = \int_{L^3} d \mathbf{ r} e^{-i \mathbf{ p} \cdot \mathbf{ r}} \Phi (\mathbf{ r}; q) ,  \ \  \mathbf{ p} = \frac{2\pi}{L} \mathbf{ n},  \ \ \mathbf{ n} \in \mathbb{Z}^3, 
\end{equation}
and the inverse Fourier transform,
\begin{equation}
 \Phi (\mathbf{ r}; q) = \frac{1}{L^3} \sum_{\mathbf{ n} \in \mathbb{Z}^3}^{ \mathbf{ p} = \frac{2\pi}{L} \mathbf{ n} }  e^{i \mathbf{ p} \cdot \mathbf{ r}} \widetilde{\Phi }(\mathbf{ p}; q)   \ .
\end{equation}
We find the momentum space representation of Eq.~(\ref{2bLuscher}),
\begin{equation}
 \widetilde{\Phi }(\mathbf{ p}; q)  = \frac{1}{q^2- \mathbf{ p}^2}  \frac{1}{L^3}  \sum_{\mathbf{ n}' \in \mathbb{Z}^3}^{ \mathbf{ p}' = \frac{2\pi}{L} \mathbf{ n}' }   \widetilde{V} (\mathbf{ p} - \mathbf{ p}')  \widetilde{\Phi }(\mathbf{ p}'; q)  ,
\end{equation}
where $ \widetilde{V} (\mathbf{ p}  )  = \int_{L^3} d \mathbf{ r} e^{-i  \mathbf{ p} \cdot \mathbf{ r}}  V(r) $. Therefore, the quantization condition may also be given by
\begin{align}
& \det \left [ \delta_{\mathbf{ p}, \mathbf{ p}'}-   \frac{1}{q^2- \mathbf{ p}^2}  \frac{1}{L^3}    \widetilde{V} (\mathbf{ p} - \mathbf{ p}')    \right ] = 0, \nonumber \\
 & \quad \quad \quad \quad ( \mathbf{ p}, \mathbf{ p}')  \in \frac{2\pi}{L} \mathbf{ n},  \ \ \mathbf{ n} \in \mathbb{Z}^3. \label{2bquantmom}
\end{align}
The momentum-space representation of the two-body quantization condition in  Eq.(\ref{2bquantmom}) is in fact consistent with the two-body quantization conditions in \cite{Doring:2011vk,Doring:2011ip} obtained  by a different method.

The   approach  described above may be generalized to multiparticle system as well \cite{Guo:2019hih}. In the following, we consider a simple $1+1$ dimensional heavy-light three-body system with one static heavy particle interacting with two light bosons. The heavy-light system resembles an atomic system  with a periodic boundary condition,  and only pair-wise $\delta$-function short-range interactions between the heavy and the light particle, and between the two light particles are considered in this work,  which may be tested by a simple $\phi^4$-type lattice model.  We remark that the short-range interactions may be described by $\delta$-function potential and its derivatives,  where $\delta$-function term may be considered as leading order contribution, and its derivatives are considered sub-leading order effects. As for the purpose of demonstrating the feasibility of our approach in present work,   only leading order contribution is considered. It turns out that the lattice model data can be fairly well described   even   without sub-leading order contributions, see Sec. \ref{data}.  The more general form of interactions, including three-body potential, will be discussed in our future publications.

\subsection{A non-relativistic heavy-light three-body system in $1+1$ dimensions}\label{nonrelform}

In finite volume with periodic boundary conditions the dynamics of a three-body system with two light scalar particles at $x_1$ and $x_2$ and one static heavy boson 
at the origin   is described by the Schr\"odinger equation,
\begin{equation}
\left [ \sigma^2 + \hat{T} - U_L (x_1) - U_L (x_2) - V_L(r) \right ] \Phi(x_1,x_2)=0, \label{3bschro}
\end{equation}
where  $\sigma^2 = 2 m E$, $ \hat{T} = \frac{\partial^2}{\partial x_1^2} +  \frac{\partial^2}{\partial x_2^2} $,  $r=x_1 - x_2$.   The potentials  $U_L (x) =   U_0 \sum_{n \in \mathbb{Z}} \delta (x + n L)$  are the short-range interactions  in the heavy-light two-body sub-systems   and $V_L( r) =   V_0 \sum_{n\in \mathbb{Z}} \delta( r+n L)$    is the interaction between the light particles. $L$ stands for the size of the one-dimensional periodic box. The wave function in finite volume satisfies periodic boundary conditions,
\begin{equation}
 \Phi(x_1 + n_1 L,x_2 + n_2 L) =  \Phi(x_1,x_2), \ \ n_{1,2} \in \mathbb{Z} \ .
\end{equation}
The integral representation of Eq.~(\ref{3bschro}) is given by a homogeneous Lippmann-Schwinger equation,
\begin{align}
& \Phi(x_1,x_2) =    \int_{-\frac{L}{2}}^{\frac{L}{2}} d x'_1 d x'_2 G_0^L (x_1-x'_1, x_2-x'_2 ; \sigma) \nonumber \\
& \times  \left [ U_0 \delta(x'_1) + U_0 \delta(x'_2) + V_0 \delta(x'_1-x'_2)  \right ] \Phi(x'_1,x'_2) ,   \label{3blipp}
 \end{align}
 or more explicitly
 \begin{align}
 &\Phi(x_1,x_2) =   U_0 \int_{-\frac{L}{2}}^{\frac{L}{2}}  d x'    G_0^L (x_1 , x_2-x' ; \sigma)  \Phi( 0,x')   \nonumber \\
 &+   U_0 \int_{-\frac{L}{2}}^{\frac{L}{2}}  d x'    G_0^L (x_1-x', x_2  ; \sigma)   \Phi(x',0)   \nonumber \\
& +V_0 \int_{-\frac{L}{2}}^{\frac{L}{2}} d x' G_0^L (x_1-x', x_2-x' ; \sigma)  \Phi(x',x'), 
\end{align}
 where the periodic  three-body Green's function $G_0^L$ satisfies 
 \begin{equation}
\left ( \sigma^2 + \hat{T}  \right )   G_0^L (x_1 , x_2  ; \sigma) = \sum_{n_1,n_2 \in \mathbb{Z}} \delta (x_1 + n_1 L) \delta (x_2 + n_2 L)\ . \label{G0Leq}
\end{equation}
The analytic expression of $G_0^L$ is given by
\begin{equation}
 G_0^L (x_1 , x_2  ; \sigma)  = \frac{1}{L^2} \sum_{n_1, n_2 \in \mathbb{Z}}^{p_{i} = \frac{2\pi}{L} n_i } \frac{e^{i ( p_1 x_1 + p_2 x_2 ) }}{\sigma^2 - p_1^2 - p_2^2} \ . \label{G0LsumHL3b}
\end{equation}
Typically, the representation of the lattice Green's function in Eq.~(\ref{G0LsumHL3b})
exhibits poor convergence, and Ewald's method has been widely used to improve it \cite{doi:10.1002/andp.19213690304}.  For the completeness of our presentation, rapidly converging representations of $G_0^L$ are given in Appendix   \ref{latsumgreen}.

Using the fact that the two light particles are identical, $ \Phi(x_1,x_2)  =  \Phi(x_2,x_1) $,  Eq.~(\ref{3blipp})
can be reduced to   coupled equations,
\begin{align}
&\Phi(x,x)  =     \int_{-\frac{L}{2}}^{\frac{L}{2}} d x'   V_0 G_0^L (x-x', x-x' ; \sigma)  \Phi(x',x')  \nonumber \\
& +    \int_{-\frac{L}{2}}^{\frac{L}{2}} d x'   2 U_0 G_0^L (x, x-x' ; \sigma)  \Phi(x',0) ,  \label{3bcoupeq1} \\
& \Phi(x,0)  =      \int_{-\frac{L}{2}}^{\frac{L}{2}} d x'   V_0 G_0^L (-x', x-x' ; \sigma)  \Phi(x',x')  \nonumber \\
& +    \int_{-\frac{L}{2}}^{\frac{L}{2}} d x'    U_0  \left [ G_0^L (0, x-x' ; \sigma) + G_0^L ( - x', x ; \sigma) \right ] \Phi(x',0) . \label{3bcoupeq2}
\end{align}
By introducing a column vector, $\xi(x) = \left [\Phi(x,x) , \Phi(x,0) \right  ]^T$,  these coupled equations may be expressed in a  simple form,
\begin{equation}
\xi(x)  = \int_{-\frac{L}{2}}^{\frac{L}{2}} d x'  \mathcal{G}^L(x,x';\sigma)  \xi(x'), \label{3bcoupmat}
\end{equation}
where $\mathcal{G}^L$ is a $2\times 2$ matrix function,
\begin{align}
 \mathcal{G}^L_{1,1}(x,x';\sigma)  & =       V_0 G_0^L (x-x', x-x' ; \sigma), \nonumber \\
  \mathcal{G}^L_{1,2}(x,x';\sigma)  & =  2  U_0 G_0^L (x, x-x' ; \sigma), \nonumber \\
    \mathcal{G}^L_{2,1}(x,x';\sigma)  & =  V_0 G_0^L (-x', x-x' ; \sigma) , \nonumber \\
      \mathcal{G}^L_{2,2}(x,x';\sigma)  & =   U_0  \left [ G_0^L (0, x-x' ; \sigma) + G_0^L ( - x', x ; \sigma) \right ] . \label{Gcornonrel}
\end{align}
The quantization condition may   be obtained by applying the variational method  \cite{Guo:2018ibd,Guo:2019hih}.  Assuming that the wave function $\xi (x)$ may be expanded in terms of some known orthonormal  basis functions  $\chi_J (x)$:    $\xi(x)  = \sum_J  c_J \chi_J (x)$, Eq.~(\ref{3bcoupmat})   leads to a secular equation of familiar form,
\begin{equation}
\sum_{J'}\left [ \delta_{J,J'} -  \mathcal{G}^L_{J,J'}(\sigma)  \right ] c_{J'} =0,
\end{equation}
where $ \mathcal{G}^L_{J,J'}(\sigma)  =  \int_{\cor{}{-}\frac{L}{2}}^{\frac{L}{2}} d x d x'  \chi^*_J (x)  \mathcal{G}^L(x,x';\sigma) \chi_{J'} (x')$.  Hence, a non-trivial solution  $\xi (x)$ exists, provided the determinant condition,
\begin{equation}
\det \left [ \delta_{J,J'} -  \mathcal{G}^L_{J,J'}(\sigma)  \right ] =0,
\end{equation}
is satisfied. 

\subsubsection{Quantization condition in coordinate representation}
The major goal of this work is to   obtain   a quantization conditions which is easy to be used to  produce the discrete energy spectra and fit data of lattice model simulations.  As far as the basis functions used in the variational method converge, the quantization conditions obtained by the variational method should not depend on the choice of basis  ultimately.  However, in reality, projecting out the matrix elements in the variational method \cite{Guo:2018ibd,Guo:2019hih} by multiple dimensional integration is still an computationally intense
task. Fortunately, by discretizing the homogeneous Eq.~(\ref{3bcoupmat}),
\begin{equation}
\xi_{  i}   = \sum_j w_j \mathcal{G}^L (x_i,x_j;\sigma)  \xi_j, \label{3bdiscret}
\end{equation}
where $\xi_i = \xi (x_i)$, and $x_j$  and $w_j$ stand for the nodes and weights   of coordinate discretization in the range of $x_j \in [-\frac{L}{2}, \frac{L}{2}]$, the $\xi_i$ can be treated as   basis functions for the variational method. Therefore, a quantization condition without any dependence of specific choice of basis functions is given by
\begin{equation}
\det \left [ \delta_{\alpha, \beta} \delta_{i,j}   -  w_j \mathcal{G}^L_{\alpha, \beta}(x_i,x_j;\sigma ) \right ] =0 \ . \label{3bquantcorr}
\end{equation}
One last remark about the quantization condition given by Eq.~(\ref{3bquantcorr})  is that due to the singularity of $G_0^L (x_1, x_2 ; \sigma)   \sim    \frac{1}{2\pi} \ln \sigma |\mathbf{ x}| $ as $|\mathbf{ x}| = \sqrt{x_1^2 + x_2^2} \sim 0$, the singular terms of the diagonal matrix elements, $\mathcal{G}^L_{\alpha, \alpha}(x_i,x_i;\sigma )$, have to be regularized. A modified subtraction quadrature method \cite{kythe2002computational} may be used for the regularization scheme of singularities, the details are presented in Appendix \ref{detsub}.

\subsubsection{Quantization condition in momentum representation}
Introducing Fourier transformation in finite volume, $\widetilde{\xi} (p) =  \int_{-\frac{L}{2}}^{\frac{L}{2}} d x e^{- i p x}  \xi(x)= \left [ \widetilde{\xi}_1 (p) , \widetilde{\xi}_2 (p)  \right ]^T$ with $p = \frac{2\pi}{L}n $  $(n \in \mathbb{Z})$, and also with the help of the identity,
\begin{equation}
\frac{1}{L} \sum_{n \in \mathbb{Z}}^{p = \frac{2\pi}{L} n} \frac{1}{q^2 - p^2} = \frac{\cot \frac{\sqrt{q^2} L}{2}}{2 \sqrt{q^2}},
\end{equation}
  an equivalent representation of Eq.~(\ref{3bcoupmat}) in momentum space may be found,
\begin{equation}
\widetilde{\xi} (p)  = \sum_{n' \in \mathbb{Z}}^{p' = \frac{2\pi}{L} n'}  \widetilde{\mathcal{G}}^L( p, p';\sigma)  \widetilde{\xi} (p')  ,
\end{equation}
where
\begin{align}
\widetilde{\mathcal{G}}^L_{1,1}( p, p';\sigma) &  =   \delta_{p, p'}   V_0 \frac{ \cot \frac{\sqrt{ \frac{\sigma^2}{2} - \frac{p^2}{4} } + \frac{p}{2}  }{2} L }{4 \sqrt{ \frac{\sigma^2}{2} - \frac{p^2}{4} }}     , \nonumber \\
\widetilde{\mathcal{G}}^L_{1,2}( p, p';\sigma) & =  \frac{  U_0}{L} \frac{2}{\sigma^2 - {p'}^2 - (p-p')^2}, \nonumber \\
\widetilde{\mathcal{G}}^L_{2,1}( p, p';\sigma) & =  \frac{  V_0}{L} \frac{1}{\sigma^2  - (p-p')^2 - p^2 }, \nonumber \\
\widetilde{\mathcal{G}}^L_{2,2}( p, p';\sigma) &  =   \delta_{p, p'}    U_0 \frac{ \cot \frac{\sqrt{  \sigma^2  - p^2 }  }{2} L }{ 2 \sqrt{ \sigma^2  - p^2  }}   \nonumber \\
&   +    \frac{  U_0}{L} \frac{1}{\sigma^2  -  {p'}^2 - p^2 }  .
\end{align}
Therefore, in momentum space, $\widetilde{\xi} (p)$ with discrete momenta, $p = \frac{2\pi}{L}n $  $(n \in \mathbb{Z})$, may be treated as variational basis functions, so, the quantization condition is accordingly given by
\begin{equation}
 \det \left [ \delta_{\alpha, \beta} \delta_{p,p'}   -   \widetilde{ \mathcal{G}}^L_{\alpha, \beta} ( p, p';\sigma)   \right ] =0  ,    \label{3bquantmom}
\end{equation}
where  $(p,p') \in \frac{2\pi }{L} n$,   $n \in \mathbb{Z} $.

\subsection{Consistency check at extreme limits}
We consider two extreme limits of the heavy-light system as a simple check.
\subsubsection{Limit of $U_0=0$}
In the limit   $U_0=0$, the two light particles decouple completely from the interaction of the static heavy boson, and the quantization condition Eq.~(\ref{3bquantmom}), hence, is reduced to a familiar form,
\begin{equation}
  \cot \frac{q + \frac{p}{2}  }{2} L =   \frac{4 q}{ V_0}     , \label{zeroU0nonrel}
\end{equation}
where $q= \sqrt{ \frac{\sigma^2}{2} - \frac{p^2}{4} } = \frac{p_1-p_2}{2}$ stands for the relative momentum of two light particles, and $p=\frac{2\pi}{L}  n$ with $n \in \mathbb{Z}$ represents the total momentum of two light particles. This is exactly what we expect for two interacting particles in a periodic box \cite{Guo:2013vsa,Guo:2016fgl,Guo:2017ism,Guo:2017crd}.

\subsubsection{Limit of $V_0=0$}
In the limit   $V_0=0$, the two light particles do not interact with each other, hence the coupled homogeneous Lippmann-Schwinger equation is reduced to a single one,
 \begin{align}
& \left [ 1 -  U_0 \frac{ \cot \frac{\sqrt{  \sigma^2  - p^2 }  }{2} L }{ 2 \sqrt{ \sigma^2  - p^2  }}  \right ] \widetilde{\xi}_2 (p) \nonumber \\
&  =\frac{  U_0}{L}  \sum_{n' \in \mathbb{Z}}^{p' = \frac{2\pi}{L} n'} \frac{1}{\sigma^2  -  {p'}^2 - p^2 }   \widetilde{\xi}_2 (p')  , \label{zeroV0}
\end{align}
where  $\widetilde{\xi}_2 (p) =  \int_{-\frac{L}{2}}^{\frac{L}{2}} d x e^{- i p x}  \Phi(x,0 )  $ with $p =\frac{2\pi}{L} n, \ \ n \in \mathbb{Z}$.  

To check consistency of Eq.~(\ref{zeroV0}), we note that for the case of zero   interaction between the two light particles, the wave function of
the two light   particles  is   simply given by the product of two single-particle wave functions, 
\begin{equation}
\Phi(x_1,x_2) = \phi(x_1, p_1) \phi (x_2, p_2)  + (x_1 \leftrightarrow x_2),  
\end{equation}
where  $\sigma^2= p_1^2 + p_2^2$, and $ \phi(x_i, p_i) = \frac{ t (p_i) }{2 p_i} \left [ e^{i p_i |x_i|}  + \frac{2 \cos p_i x_i}{e^{-i p_i L}-1} \right ]$ is the solution of the finite-volume single-particle wave function in the presence of the static heavy-particle interaction, and $t(q) = - \frac{ U_0}{2 q + i U_0}$ stands for light-heavy two-body scattering amplitude. The allowed momenta of the light particle are determined by the two-body quantization condition: $\cot \frac{p_i L}{2} = \frac{2 p_i}{U_0}$ \cite{Guo:2016fgl,Guo:2013vsa}. Therefore,  the expression of $ \widetilde{\xi}_2$ is given by 
\begin{equation}
\widetilde{\xi}_2 (p)  = \frac{ i t(p_1) i t(p_2)}{U_0} \left (    \frac{1 }{p_1^2 - p^2} +    \frac{1 }{p_2^2 - p^2} \right ) .
\end{equation}
Using the identity
\begin{align}
&  \frac{1}{L} \sum_{n' \in \mathbb{Z}}^{p' = \frac{2\pi}{L} n'} \frac{1}{\sigma^2  -  {p'}^2 - p^2 }   \frac{1 }{p_i^2 - {p'}^2} \nonumber \\
& = \frac{1}{\sigma^2- p^2- p_i^2 } \left [  \frac{ \cot \frac{p_i L}{2}  }{2 p_i    }  -   \frac{ \cot \frac{\sqrt{\sigma^2 - p^2}}{2}}{2  \sqrt{\sigma^2 -p^2}  }      \right ],
\end{align}
and also the two-body quantization condition,   we find
 \begin{align}
&\frac{  U_0}{L}  \sum_{n' \in \mathbb{Z}}^{p' = \frac{2\pi}{L} n'} \frac{1}{\sigma^2  -  {p'}^2 - p^2 }   \widetilde{\xi}_2 (p')  \left [ \frac{U_0}{ i t(p_1) i t(p_2)}  \right ] \nonumber \\
&  =  \left [ 1 - U_0 \frac{  \cot \frac{\sqrt{\sigma^2 - p^2}}{2} }{2   \sqrt{\sigma^2 -p^2}  } \right ]   \left (  \frac{ 1 }{p_1^2- p^2 } + \frac{ 1 }{p_2^2- p^2 } \right ),
\label{gleichung}
\end{align}
which is indeed consistent with Eq.~(\ref{zeroV0}).


\section{A $1+1$ dimensional  heavy-light system of lattice   model }\label{phi4sols}

In this section, we present a simple $1+1$ dimensional coupled-channel $\phi^4$-type lattice model to simulate a heavy-light three-body system which is described in Section \ref{3bdynamics}.
 The classical action of  a heavy-light bosonic system of the lattice model in two-dimensional  Euclidean space   is given by
\begin{align}
S  =   \int d^{2}x \bigg [ & \frac{1}{2}    \left (  \partial \phi  \right )^2  + \frac{1}{2} \mu^{2}  \phi^{2} + \frac{g_{\phi \phi} }{4!} \phi^{4}  \nonumber \\
&+ g_{\sigma \phi} \delta_{x_1,0} \phi^2 (x_0,0)  \sigma^2 (x_0,0)   \bigg ]  , \label{phi4action}
 \end{align}
where $x=(x_{0},x_{1})$ are temporal and spatial coordinates in two-dimensional Euclidean space,  respectively. The light scalar particles are represented by the $\phi (x_0,x_1)$ field, and the $\sigma(x_0,0)$ field denotes the static heavy particle  at the origin. The interaction between the light particles   is simply described by a $\phi^4$ model,  and the interaction between light and   heavy particles is described by  the coupling between the $\sigma$ and $\phi$ fields at the origin: $x_1=0$.   No kinetic term is needed for a static heavy $\sigma$ particle.

In infinite space with an open boundary condition, the heavy-light system given in Eq.~(\ref{phi4action})  is  equivalent to  a non-relativistic one-dimensional multiparticle  interacting system with $N$-light-scalar  plus one static heavy boson. Interactions between both light-light and heavy-light particles  are described by short-range  pair-wise  $\delta$-function   potentials, 
\begin{equation}
2 m H  =\sum_{i=1}^N \left [   -    \frac{\partial^2}{\partial x_{1,i}^2} + U_0 \delta( x_{1,i} ) \right ] + V_0  \sum_{i<j} \delta(x_{1,i} - x_{1,j})   , \label{HamiltNbody} 
\end{equation}
where $x_{1,i}$ refers to the spatial position of the $i$-th light scalar particle, and $m$ stands for the mass of the light   bosons. The coupling strengths  of $\delta$-function potential  among light-light and heavy-light particles are denoted by $V_0$ and $U_0$, respectively. When the periodic boundary condition is considered, the lattice model designed in Eq.~(\ref{phi4action})    can be used to simulate a finite-volume heavy-light multiparticle system that is the simple model we  study in this work.

\subsection{ The lattice   model action of heavy-light system }\label{lattmodel}
The lattice action that describes heavy-light particles living in a discrete  $1+1$ rectangle  lattice is obtained by replacing the continuous derivative with discrete difference  in Eq.~(\ref{phi4action}):  $\partial  \phi(x) \rightarrow  \phi(x+\hat{n}) - \phi(x)$,  where $\hat{n}$ denotes the unit vector in direction $x_i$. Introducing       parameters  according to $\mu^2 = \frac{1-2 \lambda_{\phi\phi}}{\kappa}-8$, $g_{\phi \phi}=\frac{6 \lambda_{\phi \phi}}{\kappa^2}$, and $g_{\sigma \phi} = \frac{\lambda_{\sigma \phi}}{2 \kappa}$, and    rescaling the $\phi$ field by $\phi \rightarrow \sqrt{2\kappa } \phi$,  we find
\begin{align}\label{action}
& S_{lat}   =\sum_{x}    \bigg [ - 2   \kappa  \sum_{ \hat{n}} \phi(x) \phi(x+\hat{n}) + (1-2 \lambda_{\phi \phi})    \phi^2(x)   \bigg ] \nonumber \\
& + \sum_x \bigg [ \lambda_{\phi \phi} \phi^4(x)  +\lambda_{\sigma \phi}   \delta_{x_1,0} \phi^2 (x_0,0)  \sigma^2 (x_0,0)  \bigg ]  ,
\end{align}
where $x=(x_{0},x_{1})$ now refers to the discrete coordinates of the  Euclidean $T\times L$ rectangle   lattice sites.

\subsection{ Monte Carlo updating algorithm for heavy-light system}
  A combination of Hybrid Monte Carlo updating algorithm \cite{Duane:1987de,Duane:1986iw}  for the $\phi$ field and the standard  Metropolis–Hastings updating algorithm \cite{Hastings1970,Metropolis1953} for the $\sigma$ field   is adopted in our   simulation. The $\phi$ and $\sigma$ fields are updated  alternately.  For  updating the $\phi$ field with the Hybrid Monte Carlo algorithm, an auxiliary Hamiltonian and a  fictitious conjugate momentum of the $\phi $ field,  the $\pi  $ field,  is introduced,
\begin{align}
H_{lat} = \frac{1}{2} \sum_x \pi^2(x)  + S_{lat} \ . \label{hamiltonian}
\end{align}
The auxiliary Hamiltonian in Eq.~(\ref{hamiltonian}) defines  the classical evolution of both $\pi$ and $\phi$ fields  over a  fictitious time  within an interval $  [0, \tau]$:
\begin{align}
&   \phi  (\tau ) =  \phi  (0 ) + \int_0^{\tau} d \tau' \pi  (\tau' ),   \nonumber \\
&   \pi (\tau ) = \pi  (0 )  - \int_0^{\tau} d \tau'  \frac{\partial S_{lat}(\phi (\tau' ) )}{ \partial \phi (\tau' )} \ . \label{eqofmotion}
\end{align}
The trajectory of $(\phi,\pi)$ over the  time interval $  [0, \tau]$ is determined by the solutions of  the equations  of motion in Eq.~(\ref{eqofmotion}).
The  pair of $(\phi, \pi)$ fields and $\sigma$ field  are updated alternately for each sweep over  entire lattice: 

\begin{itemize}

\item Updating  the pair $(\phi, \pi )$ with the standard Hybrid Monte Carlo algorithm:

 (i) the trajectory begins with  a random  distribution of  fields $(\phi  , \pi  )$ at initial fictitious time. The initial conjugate momenta, $\pi (0)$, are generated according to the Gaussian probability distribution: \mbox{$P(\pi ) \propto e^{- \frac{\pi^2}{2}}$}. 
 
 (ii)      $(\phi , \pi )$ are evolved    over the trajectory up to a  time $\tau$ according to  equations of motion  in Eq.~(\ref{eqofmotion}).  The equations of motion   are solved  by the leapfrog method  \cite{Duane:1986iw}. The  $(\phi,\pi)$ fields evolve along a trajectory with a fixed length \mbox{$\tau=8$} over $100$ discrete steps.  
 
  (iii)   the proposed new fields, $(\phi  , \pi  ) $, are accepted with probability:  \mbox{$ P_{acc} = \mbox{Min} \left [ 1 , e^{- \triangle H_{lat} }\right ]$}, where  \mbox{$ \triangle H_{lat} = H_{lat}(\tau) - H_{lat}(0)$}.

\item Updating the  static heavy  $\sigma (x_0, 0)$ field  with the standard Metropolis–Hastings  algorithm \cite{Hastings1970,Metropolis1953}.
 
\end{itemize}

In our  simulations,  the lattice model parameters are chosen  as:  $\kappa =0.1275$,   $\lambda_{\phi \phi}  = 0.02$ and  $\lambda_{\sigma \phi}  = 0.007$.  The temporal extent of the lattice is fixed at $T=100$,  and the   spatial extent of lattice, $L$, varies from $10$ up to $55$ with an increment of $5$.   Two million measurements are generated for   each  lattice size.


\subsection{Operator construction and particle spectra}\label{spectra}
In this section,   some details  regarding the construction of multi-particle operators    and  results for the heavy-light particle  spectra are presented.

  \begin{figure}
\begin{center}
\includegraphics[width=0.49\textwidth]{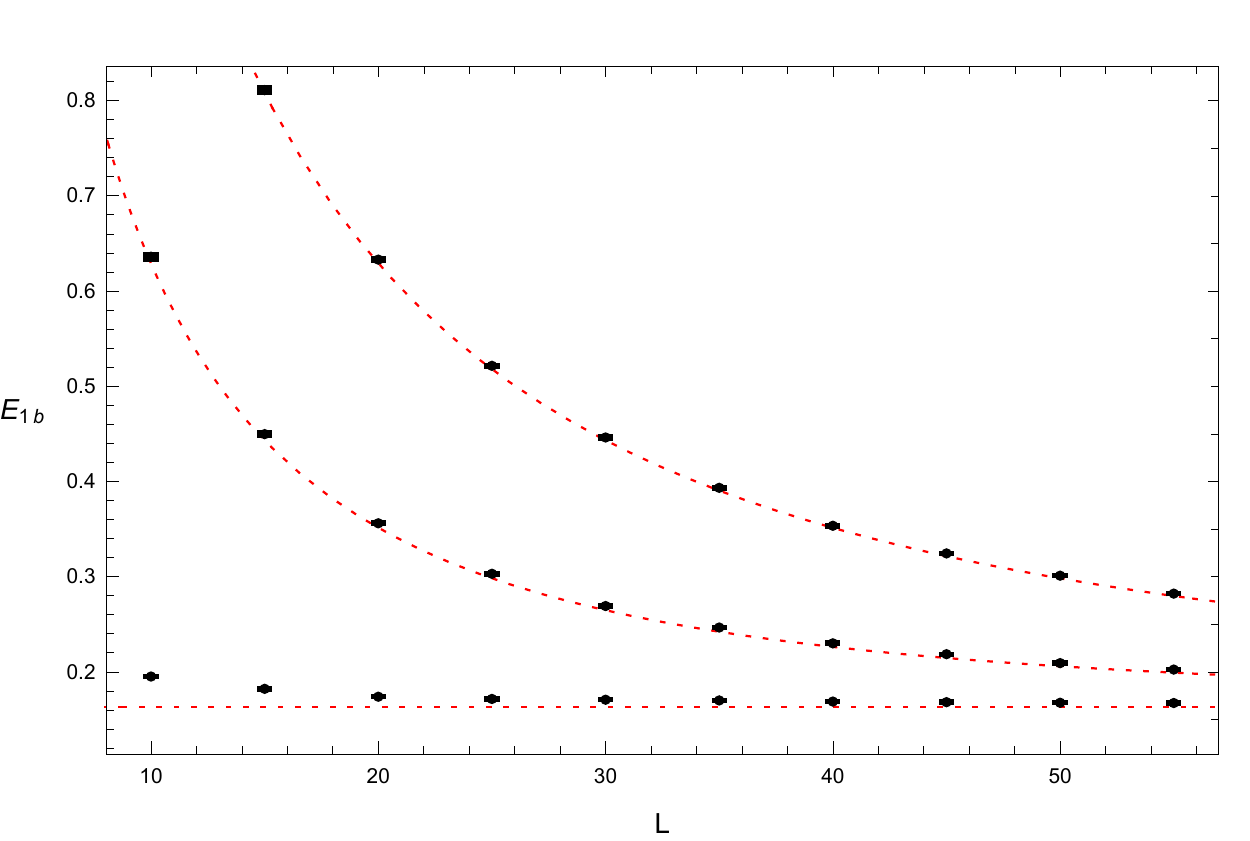}  
\caption{ Plot of single light-particle spectra in the presence of the static heavy-boson potential vs. free single-particle energy levels (red dashed curve): $E^{free}_{1b,n} (L)=  \cosh^{-1} \left ( \cosh m +1 - \cos \frac{2\pi}{L} n  \right )$ with $m=0.163$ and $n=0,1,2$. \label{E1bplot}}
\end{center}
\end{figure}

  \begin{figure}
\begin{center}
\includegraphics[width=0.45\textwidth]{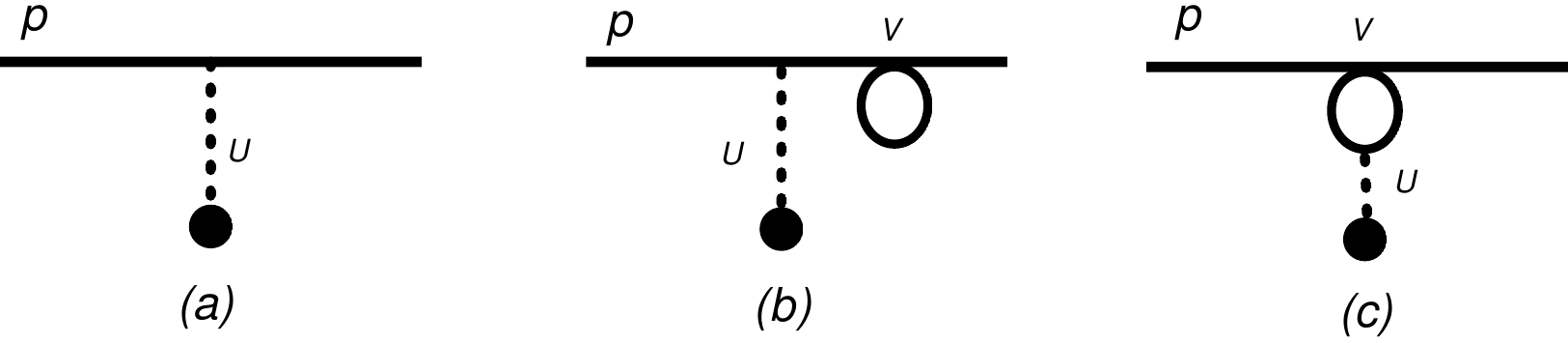}  
\caption{ Lowest few orders diagrammatic representation of heavy-light two-body interactions from point of view of perturbation theory. The solid black circles stand for the static heavy particle and the light particle is represented by the solid lines. The contact interaction between light and heavy particles is denoted by a dotted line. The intersection of two solid lines represents the contact interaction between the two light particles.  $U$ and $V$ represent the bare coupling strengths of heavy-light and light-light particles respectively. \label{oneparticleplot}}
\end{center}
\end{figure}


\subsubsection{The spectra of one light scalar in the presence of a static heavy particle potential}\label{onespectra}

In the presence of the static heavy-boson potential, the spectra of  the light scalar particle interacting with the heavy boson    are extracted from the exponential decay of the correlation functions
\begin{equation}
C_{ 1b, n}(x_{0}) = \langle   \widetilde{\phi}^*_{n}(x_{0}) \widetilde{\phi}_{n}(0) \rangle \propto e^{- E_{1b, n}  x_{0}} \ ,
\end{equation}
where  the light particle propagator, $\widetilde{\phi}_{n}(x_{0}) $, is defined by
\begin{equation}
\widetilde{\phi}_{n}(x_{0}) =\frac{1}{L} \sum_{x_{1}} \phi(x) e^{i x_{1} \frac{2\pi}{L} n },   \ \  n  \in \mathbb{Z} \ .
\end{equation}
Multiple $E_{1b, n}$ are extracted for each lattice size, see Fig.~\ref{E1bplot}.  Because of the interaction with the static heavy-boson potential,  the energy spectra   $E_{1b, n}$ can no longer be described by a simple free-particle dispersion relation with the light particle mass as a single free parameter, i.e., they are no longer of the form   $E^{free}_{1b,n} (L) =  \cosh^{-1} \left ( \cosh m +1 - \cos \frac{2\pi}{L} n  \right )$ \cite{Gattringer:1992np}, as indicated in  Fig.~\ref{E1bplot}. 
Instead, the $E_{1b, n}$  depend on two parameters:  the renormalized particle mass,
  and the  renormalized heavy-light interaction strength. The diagrammatic representations of the heavy-light two-body interactions are shown in Fig.~\ref{oneparticleplot}.

  \begin{figure}
\begin{center}
\includegraphics[width=0.49\textwidth]{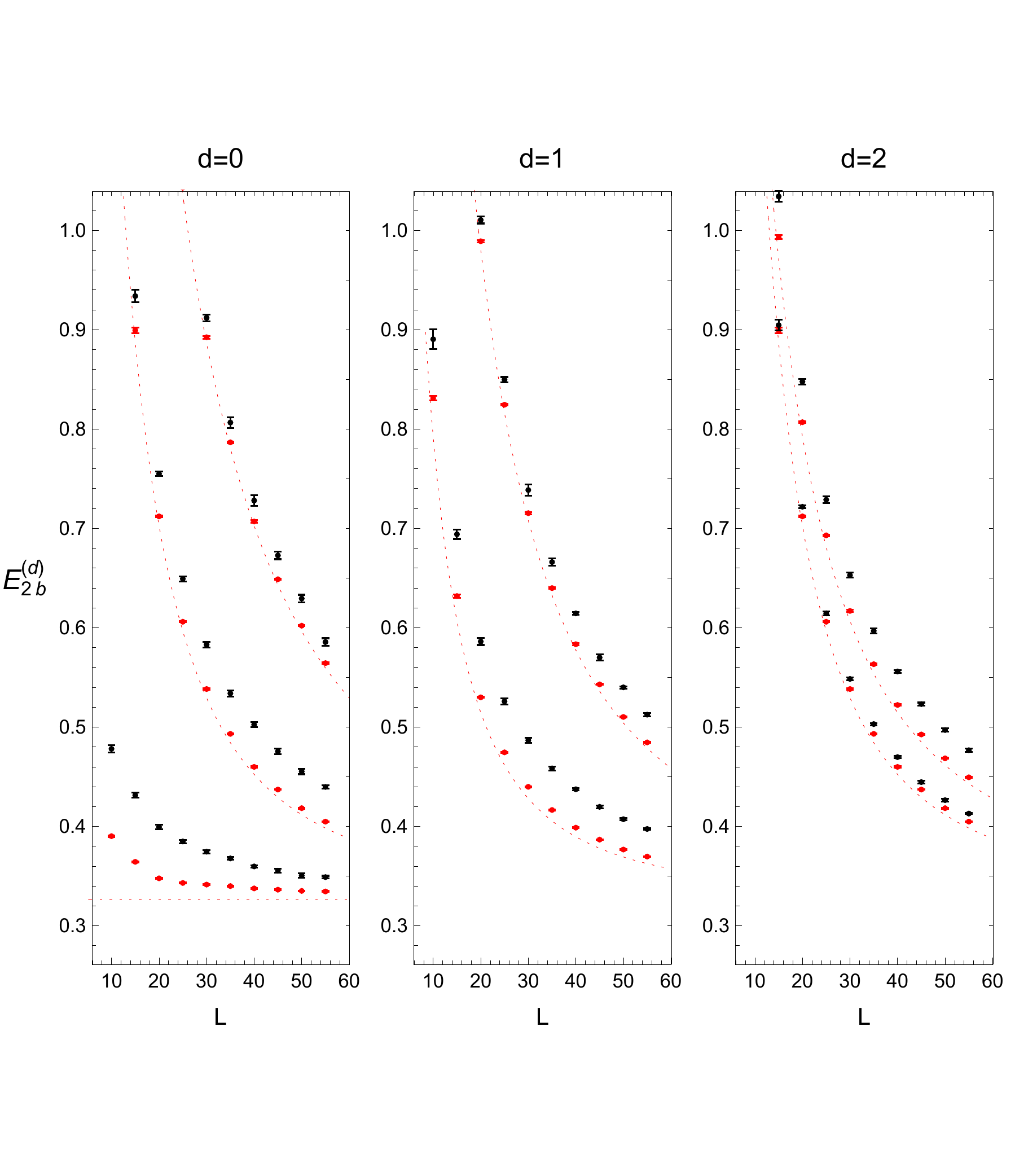}  
\caption{ Plot of  light two-particle spectra in the presence of a static heavy-boson potential (black) vs. energy levels of $E_{1b, n_1}+E_{1b,n_2}$ (red), where $E_{1b,n}$  are the single-particle energy spectra presented in Fig.\ref{E1bplot}. The red dotted curves represent spectra of two free light particles: $E^{free}_{2b} (L)= \sum_{i=1,2} \cosh^{-1} \left ( \cosh m +1 - \cos \frac{2\pi}{L} n_i  \right )$ with $m=0.163$. 
\label{E2bplot}}
\end{center}
\end{figure}

\subsubsection{The spectra of two light scalars in the presence of a static heavy particle potential}\label{twospectra}           
 
\begin{figure}
\begin{center}
\includegraphics[width=0.48\textwidth]{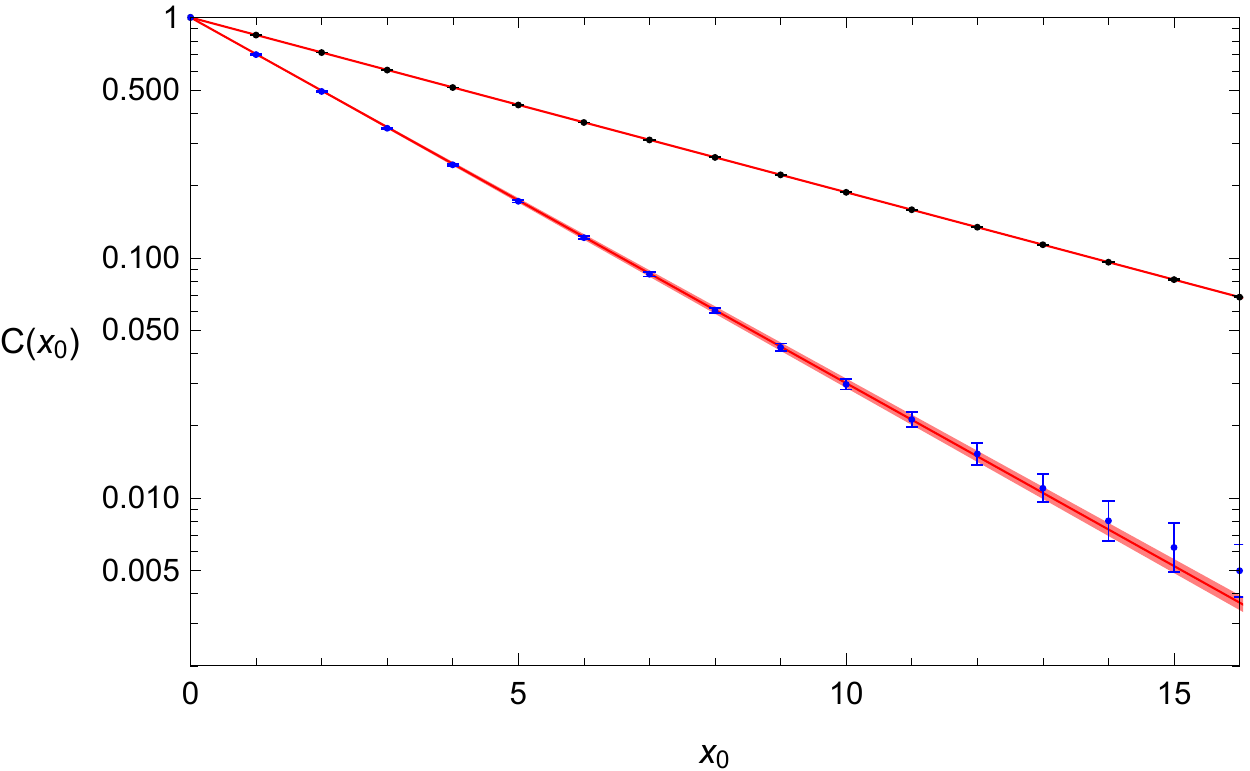}  
\caption{ Correlation functions  for one light particle (black) and two light particles (blue) at $L=50$ and $d=0$, and corresponding fitting curves (red band).  \label{corrplot}}
\end{center}
\end{figure}

\begin{figure}
\begin{center} 
\includegraphics[width=0.48\textwidth]{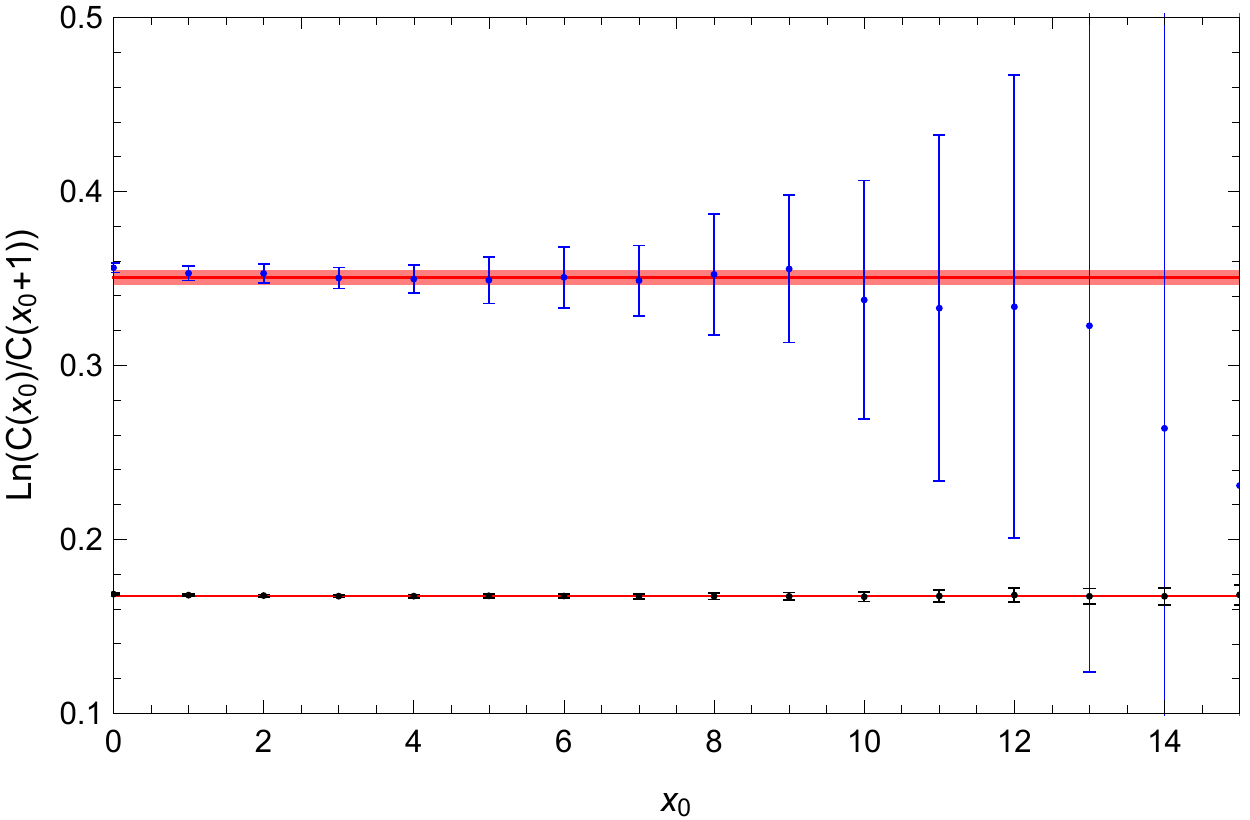}  
\caption{ Effective mass plots, $\ln \frac{C(x_{0})}{C(x_{0}+1)} $, for one light particle (black) and  two particles (blue)  at $L=50$ and $d=0$, and corresponding fitting curves (red band). \label{effmassplot}}
\end{center}
\end{figure}

The matrix element  of the correlation function of the two light particles interacting with a static heavy boson  is defined by
\begin{equation}
C^{(d)}_{2b, (i, j)} (x_{0}) = \langle   \left [  O^{(d)*}_{2b, i}(x_{0})  - \delta_{d,0} O^{(d)*}_{2b, i}(x_{0}+1) \right ] O^{(d)}_{2b, j}(0)    \rangle ,
\end{equation}
where  $d\in \mathbb{Z}$ is related to the center of mass momentum of the two light particles by $P= \frac{2\pi}{L} d$.  The disconnected contribution needs to be subtracted in the CM frame $(d=0)$.  Typically, three or four  two-light-particle operators  are used in our simulation for $d=0,1,2$,
\begin{align}
 O^{(0)}_{2b}(x_{0}) &= \widetilde{\phi}_{n}(x_{0})   \widetilde{\phi}_{-n}(x_{0}) , \ \ \ \ n =0,1,2,3, \nonumber \\
 O^{(1)}_{2b}(x_{0}) &= \widetilde{\phi}_{n}(x_{0})   \widetilde{\phi}_{1-n}(x_{0}) , \ \ \ \ n =1,2,3,  \nonumber \\
 O^{(2)}_{2b}(x_{0}) &= \widetilde{\phi}_{1}(x_{0})   \widetilde{\phi}_{1}(x_{0}), \  \   \widetilde{\phi}_{n}(x_{0})   \widetilde{\phi}_{2-n}(x_{0}) ,  \  \   n =2,3.
\end{align}
 The spectral decompositions of the correlation function matrices are   given by  
\begin{equation}
C^{(d)}_{2b,( i,  j)} (x_{0}) = \sum_{n} v^{(d,n) *}_{2b, i}v^{(d,n) }_{2b, j} e^{- E_{2b, n}^{(d)}  x_{0}},
\end{equation}
where $v^{(d,n) }_{2b, i} =\langle n  |O^{(d)}_{2b, i}(0) |0\rangle $,  and $n$ labels the $n$-th energy eigenstate $E_{2b, n}^{(d)}$.  In order to extract excited energy states,  a generalized eigenvalue method  \cite{Luscher:1990ck,Blossier:2009kd}   is  used,
\begin{equation}
C^{(d)}_{2b}(x_{0}) \varphi_{2b,n} = \lambda^{(d)}_{2b, n} (x_{0}, \bar{x}_{0}) C^{(d)}_{2b}(\bar{x}_{0}) \varphi_{2b, n} \ ,
\end{equation}
where    $\bar{x}_{0}$ is a small reference time that  is set to zero in this work.  A simple form of \mbox{$ \lambda^{(d)}_{2b, n} (x_{0},  0) = e^{-   E_{2b, n }^{(d)} x_{0} } $} is used in the data fitting for $x_0 \in [0,8]$  as no contamination from high excited states is observed.  The energy spectra of two light particles  for various lattice sizes and $d$ are presented in Fig.~\ref{E2bplot}.   Examples of both one light particle and  two light particles correlation functions,      $C(x_{0},  0) $, and effective masses, $\ln \left [ C (x_{0},  0) / C (x_{0}+1,  0) \right ] $ are given in Fig.~\ref{corrplot} and Fig.~\ref{effmassplot},  respectively.

  \begin{figure}
\begin{center}
\includegraphics[width=0.45\textwidth]{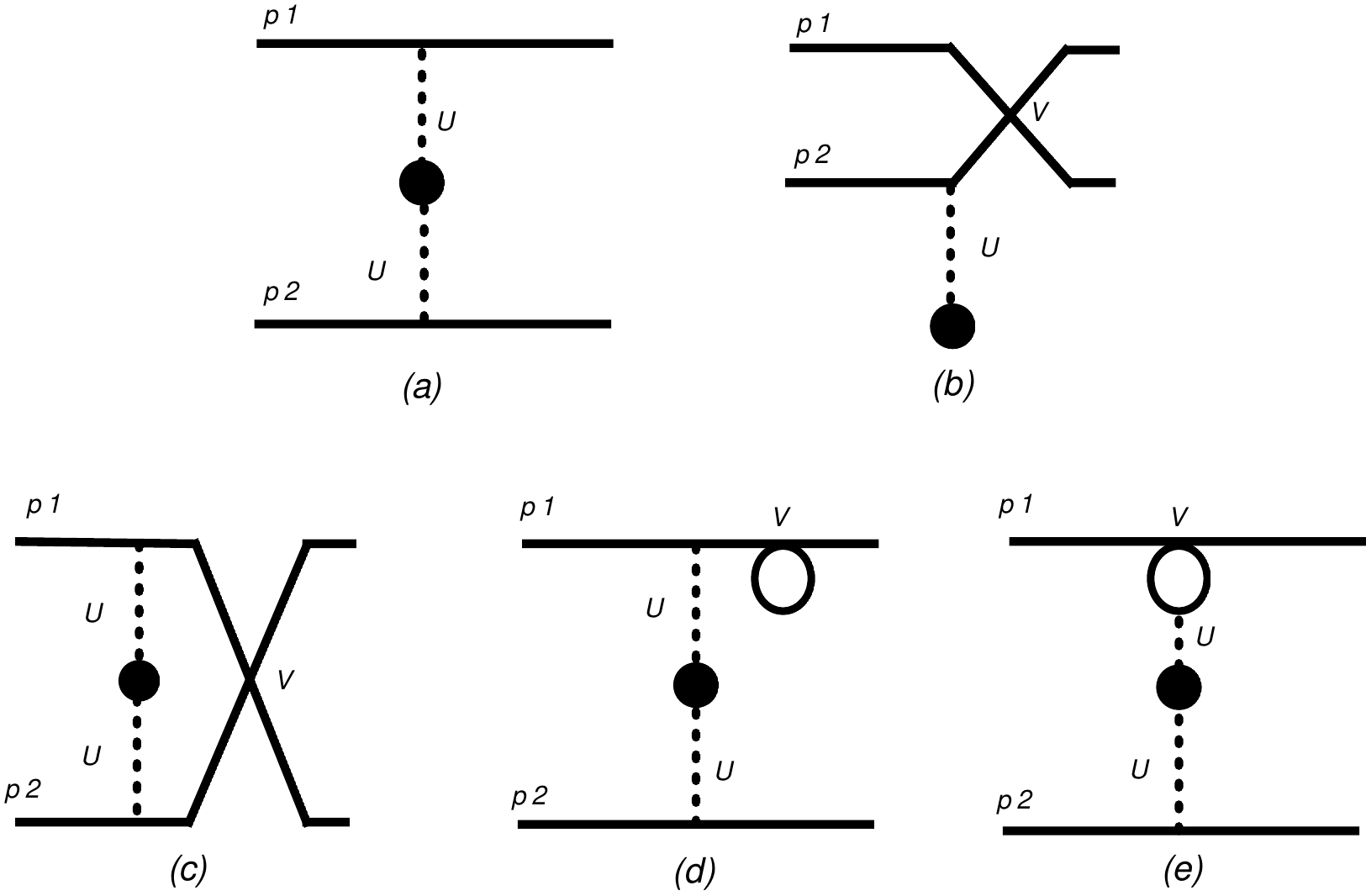}  
\caption{ Diagrammatic representation of connected heavy-light three-body interactions. For labeling see Fig.~\ref{oneparticleplot}.\label{twoparticleplot}}
\end{center}
\end{figure}

 Due to the presence of the static boson potential, the spectra $E^{(d)}_{2b, n}$ are determined by three parameters:  light particle mass, $m$,   heavy-light interaction strength, $U_0$, and light-light interaction strength, $V_0$. The     diagrammatic representation of connected heavy-light three-body interactions is shown  in Fig.~\ref{twoparticleplot}.   The sum of two one-light-particle energy spectra from the previous subsection \ref{onespectra}, $E_{1b,n_1} + E_{1b,n_2}$,   are also presented in Fig.~\ref{E2bplot}, see red error bars, which represent the results of two light particles spectra  in the absence of the interaction between the two light particles. The difference between full two light particles spectra, $E^{(d)}_{2b,n}$, and sum of two  one-light-particle energy spectra, $E_{1b,n_1} + E_{1b,n_2}$, indicates the energy shift due to the non-zero interaction between the two light particles.   One interesting observation is that  in the limit $V_0=0$,  $E^{(d=0)}_{2b}=E_{1b} (p_1) + E_{1b} (-p_1)$ (see second red energy level on left panel in Fig.~\ref{E2bplot}) and $E^{(d=2)}_{2b}=E_{1b}(p_1) + E_{1b} (p_1)$ (the first red energy level on right panel in Fig.~\ref{E2bplot})  are degenerate. However, the degeneracy is lifted because of non-zero interactions between the two light particles, see the full two light particle spectra   in second level (black error bars) on left panel and first level on right panel in Fig.~\ref{E2bplot}. We also remark that although we still use $P=\frac{2\pi}{L} d$ with $d=0,1,2$ to label two-light particles spectra, we do have to keep in mind that because of the presence of the static heavy boson, the total momentum of the two light particles, $p_1+p_2 $, is not a conserved quantity, and $p_1+p_2 \neq P=\frac{2\pi}{L} d$.


\section{Data analysis  }\label{data}

As demonstrated in Section \ref{3bdynamics},   although the finite-volume   wave function given in Eq.~(\ref{3blipp})  has no unique solution, the energy spectra of the heavy-light multiple-body system in finite volume  are uniquely determined by the multiparticle interaction and the periodic lattice structure. The quantization conditions which produce discrete energy spectra in Eq.~(\ref{3bquantcorr}) and Eq.~(\ref{3bquantmom})  do not depend on the specific  form of the finite-volume wave function, nor any particular choice of basis functions in the variational approach~\cite{Guo:2018ibd,Guo:2019hih}. 
 As demonstrated in Section \ref{3bdynamics},  quantization conditions may be constructed in such the way that   energy spectra are given   by   parameterization of potentials.
Although in QCD   multiparticle systems  in general exhibit complicated dynamics, for some simple systems at low energies,  the dynamics    may be determined by a few parameters,  \cite{Hammer:2017uqm,Hammer:2017kms}, such as the  particle masses and the interaction strength. Therefore, the main tasks of this work are   to determine these fundamental parameters   by the using quantization conditions to fit energy spectra of lattice model simulations. In this particular lattice model, there are three free parameters: the light particle mass, $m$,  and the coupling strengths of the $\delta$-potential, $U_0$ between the  heavy and the light particles, and $V_0$, between the  light particles. The light-particle mass, $m$, and coupling strength, $U_0$, may be extracted from spectra of one light particle in the presence of the static heavy boson. The coupling strength, $V_0$,  may then be  determined by studying the spectra of two light scalars in the presence of the static heavy boson.

 At this point we face two major obstacles: (1) The lattice model presented in Section \ref{phi4sols} represents a  relativistic model, and the quantization conditions given by Eq.~(\ref{3bquantcorr}) and Eq.~(\ref{3bquantmom}) are based on a non-relativistic framework. Hence, though  that framework presents a clean and simple  illustration of how the quantization conditions of three-body systems in a finite volume arise, the non-relativistic nature of the energy-momentum dispersion relation may not be capable to describe the lattice spectra;  (2) since the simulation of the lattice model is done in discrete rather than continuous spacetime, there is  the effect of finite  lattice.  Here, we take the discrete space into account but simply set the spacing to one. 
In the two-body sector,  these two challenges may  be remedied simply by adopting lattice   dispersion relation \cite{Guo:2018xbv},
\begin{equation}
 E_{1b}(L)  =   \cosh^{-1} \left ( \cosh m +1 - \cos p \right ),
\end{equation}
where momentum of the light-particle, $p$, may be determined by the  quantization condition
\begin{equation}
 \cot \frac{p L}{2} =  \frac{ 2p }{ U_0} \ .
\end{equation}
In the three-body sector,   one could be tempted to proceed similarly by using  a dispersion relation such as \cite{Guo:2018xbv}
\begin{equation}
 E^{(d)}_{2b}(L) = \sum_{i=1}^2 \cosh^{-1} \left ( \cosh m +1 - \cos p_i \right ) \ .   
\end{equation}
Unfortunately, the currently proposed approach is  limited to determine $\sigma^2 = p_1^2+p_2^2$ as apparent in the non-relativistic  quantization conditions,  Eq.~(\ref{3bquantcorr}), or Eq.~(\ref{3bquantmom}).   For these reasons, it may make more sense to find a relativistic formalism which is able to incorporate   the  presence of discrete space on the lattice. The relativistic framework may be achieved by replacing the non-relativistic Green's function in the Lippmann-Schwinger equation by a relativistic  one. This simple prescription  may be justified by the reduction of the relativistic Bethe-Salpeter equation to a relativistic Schr\"odinger equation under the assumption of an "instantaneous kernel function". 
 Of course, this reduction serves only to include relativistic kinematics and is sufficient for the purpose of mapping out finite-volume effects.
The details of the reduction procedure are presented in Appendix \ref{reductionBS}.
 
 
\subsection{Quantization condition in a discrete finite box }
In order to  take account of both the relativistic dynamics and the effect caused  by discretized Euclidean space-time, we reformulate the continuous-space  finite-volume formalism of the heavy-light system presented in Section \ref{3bdynamics} to  a relativistic finite-volume formalism defined  in discrete space. Therefore, the  particle coordinates are now  defined only on discrete integer points in a finite box of size $L$, $x \in [0, 1, \cdots, L-1]$, where we have assumed a lattice spacing  $a=1$.   The Fourier transform of a periodic function, $f(x+n L) = f(x)$ is thus defined by
 \begin{equation}
\widetilde{ f}(p) = \sum_{x \in [0,  L-1]}  e^{-i p x} f(x),  \label{fourier}
 \end{equation}
 where allowed lattice momenta are $p = \frac{2\pi}{L} n$, $n \in [0,1,\cdots, L-1]$, and $ \widetilde{ f}(p)$ is also a periodic function that satisfies $\widetilde{ f}(p+  2 \pi n ) = \widetilde{ f}(p)  $, see Ref.~\cite{Montvay:1994cy}. The  inverse Fourier transform into the discrete finite lattice is   given by
  \begin{equation}
f(x) = \frac{1}{L}\sum_{n \in [0,  L-1]}^{p = \frac{2\pi}{L} n }    e^{i p x} \widetilde{ f}(p) \ . \label{invfourier}
 \end{equation}
Hence, the continuous dispersion relation, \mbox{$E_p^2 = p^2 +m^2$}, is replaced by $(2 \sinh \frac{E_p}{2})^2 = (2\sin \frac{p}{2})^2  + (2 \sinh \frac{m}{2})^2 $ due to the discrete space-time in a finite box. In another word, we may use the following corresponding  relations between continuous and discrete lattice: $E_p \leftrightarrow 2 \sinh \frac{E_p}{2}$, $p \leftrightarrow 2\sin \frac{p}{2}$, $m \leftrightarrow 2 \sinh \frac{m}{2}$, and the on-shell energy-momentum relation  in the discrete lattice is now given by 
\begin{equation}
2 \sinh \frac{E_p}{2} = \sqrt{2 \cosh m -2 \cos p } \ .
\end{equation}

  \begin{figure}
\begin{center}
\includegraphics[width=0.49\textwidth]{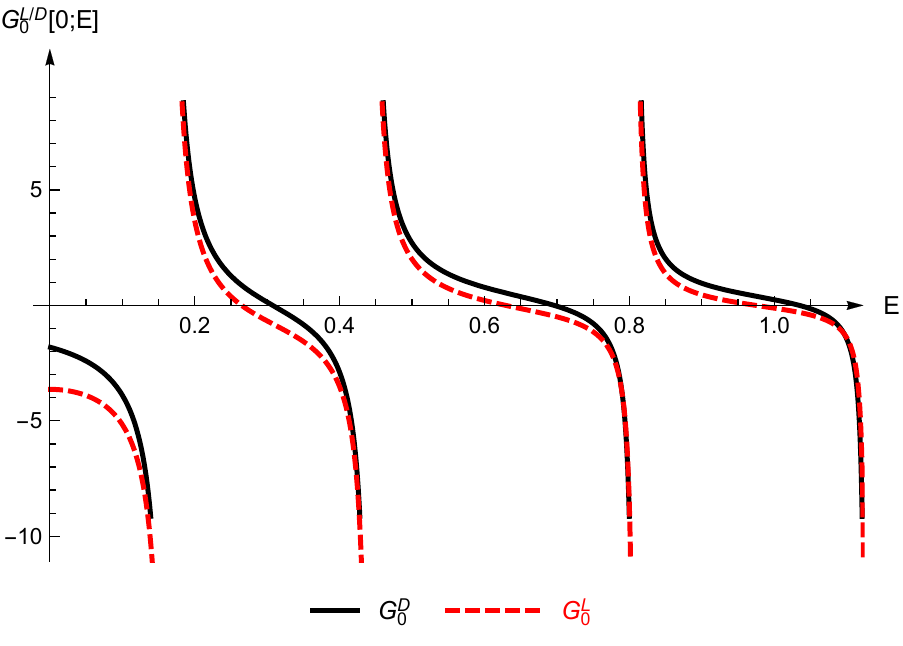}  
\caption{ Plot of two-body Green's function,  $ G_0^D (0 ; E) $ (solid black)  defined in Eq.~(\ref{G2bLat}), as function of total energy $E$   vs. non-relativistic Green's function $G_0^L (0, p) = \frac{\cot \frac{p L}{2}}{2 p}$ (dashed red) as function of energy by using the lattice energy-momentum dispersion relation, $\cosh E = \cosh m + 1 - \cos p$.     The lattice size and mass of the light particle are: $L=15$ and $m=0.163$. \label{Glatplot}}
\end{center}
\end{figure}

\subsubsection{Two-body quantization condition}
For the two-body heavy-light system, the relativistic Lippmann-Schwinger equation in discrete space may be defined by
\begin{equation}
 \Phi^D(x) =    \sum_{x' \in [0, L-1]}  G_0^D (x-x' ; E)   U_0 \delta_{x',0}  \Phi^D(x' ) ,   
 \end{equation}
 where $x \in [0, L-1]$ is defined in a  discrete finite box. The relativistic Green's function is given by
 \begin{equation}
 G_0^D (x ; E) = \frac{1}{L} \sum_{n \in [0,L-1]}^{p=\frac{2\pi}{L} n} \frac{1}{4 \sinh \frac{E_p}{2} } \frac{e^{i p x}}{ 2 \sinh \frac{E}{2} - 2 \sinh \frac{E_p}{2} }, \label{G2bLat}
 \end{equation}
 where $2 \sinh \frac{E_p}{2} =\sqrt{2 \cosh m -2 \cos p } $. In the  continuum limit, it then reduces to a relativistic-type Green's function,
  \begin{equation}
 G_0^D (x ; E) \rightarrow \frac{1}{L} \sum_{n \in \mathbb{Z}}^{p=\frac{2\pi}{L} n } \frac{1}{ 2  \sqrt{p^2+m^2}} \frac{e^{i p x}}{ E -  \sqrt{p^2+m^2} }. 
 \end{equation}
 For the $\delta$-function potential, the quantization condition for the two-body system is simply given by
 \begin{equation}
1 =   U_0   G_0^D (0 ; E)      . \label{2bquantlat}
 \end{equation}
 The difference between $G_0^D (0 ; E)$ defined in Eq.~(\ref{G2bLat}) and its non-relativistic counterpart, $G_0^L (0, p) = \frac{\cot \frac{p L}{2}}{2 p}$,  combined  with the lattice energy-momentum dispersion relation, $\cosh E = \cosh m + 1 - \cos p$, is shown in Fig.\ref{Glatplot}. 
 
\begin{figure}
\begin{center}
\includegraphics[width=0.49\textwidth]{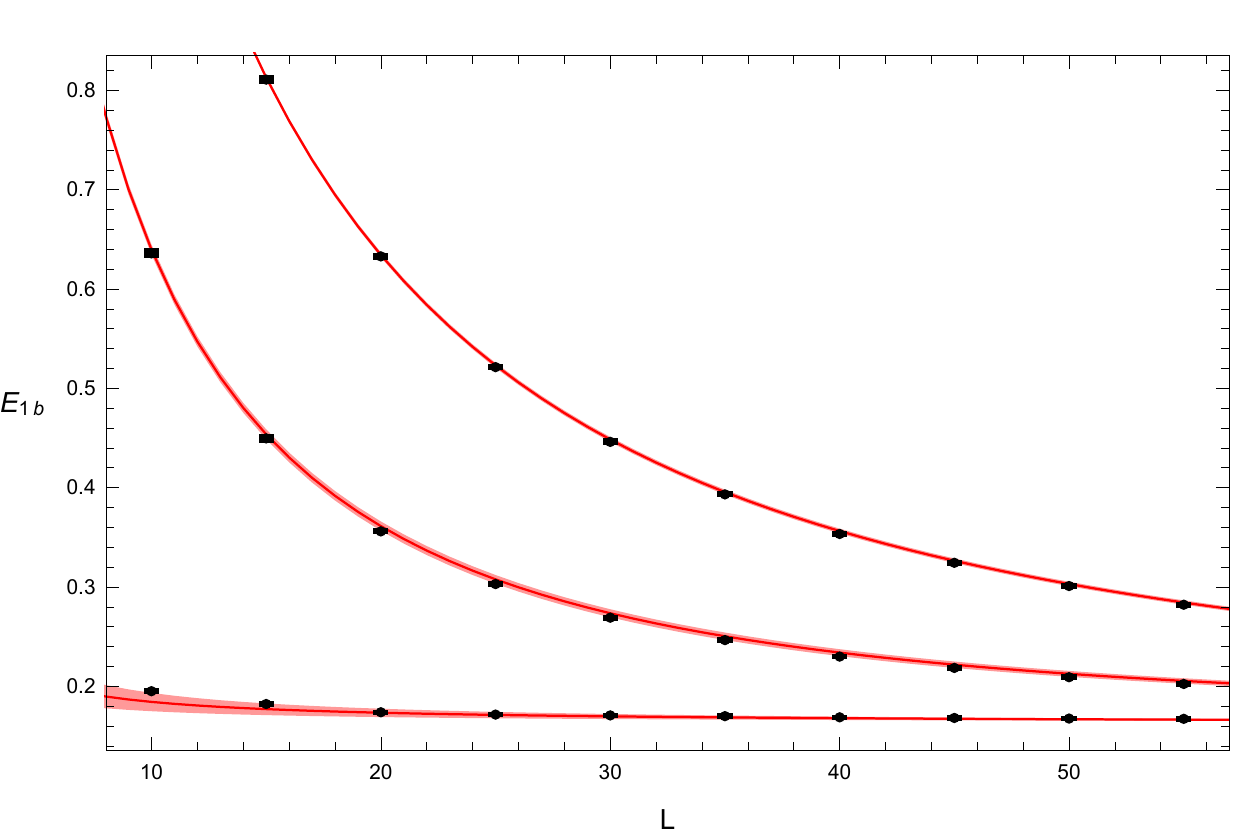}  
\caption{ Plot of single light particle spectra in presence of static heavy boson potential vs. fitting result (red  band)  by using the  two-body quantization condition of Eq.~(\ref{2bquantlat}).   \label{E1bfitplot}}
\end{center}
\end{figure}

 We test the relativistic framework by 
 using the two-body quantization condition, Eq.~(\ref{2bquantlat}),   to fit  the single-light-particle spectrum  by using  the light particle mass, $m$, and the heavy-light short-range potential coupling strength, $U_0$, as   fitting parameters. The  results are show in Fig.~\ref{E1bfitplot}, and the fitting parameters are: $m=0.163 \pm 0.001$ and $U_0 = 0.07 \pm 0.03$.

\subsubsection{Three-body quantization condition}

For the three-body system, we   need to consider a lattice version of the Lippmann-Schwinger equation defined  on a discrete lattice,
\begin{align}
&  \Phi^D (x_1,x_2)  = \sum_{x'_1,x'_2  \in   [0,L-1]}    \nonumber \\
 &\times   \bigg [  G_{(1)}^D (x_1 - x'_1 , x_2 - x'_2  ; E)    U_0 \delta_{x'_1,0}      \nonumber \\
& \quad  + G_{(2)}^D (x_1 - x'_1 , x_2 - x'_2  ; E)   U_0  \delta_{x'_2,0}    \nonumber \\
  & \quad + G_0^D (x_1 - x'_1 , x_2 - x'_2  ; E)    V_0 \delta_{ x'_1 , x'_2 }   \bigg ] \Phi^D (x'_1,x'_2)  ,
\end{align}
where  $(x_1,x_2) \in [0, L-1]$, and the Green's function's are defined by
\begin{align}
&  G_{(i)}^D (x_1 , x_2  ; E)  
  = \frac{1}{L^2} \sum_{n_1, n_2 \in [0,L-1]}^{p_{i} = \frac{2\pi}{L} n_i }  \frac{1 }{    4\sinh \frac{E_i}{2}} \nonumber \\
 &  \quad\quad  \times  \frac{e^{i ( p_1 x_1 + p_2 x_2 ) }}{ 2 \sinh \frac{E}{2} - ( 2 \sinh \frac{E_1}{2}+ 2 \sinh \frac{E_2}{2})  } , \nonumber \\
 &  G_0^D (x_1 , x_2  ; E)  
  = \frac{1}{L^2} \sum_{n_1, n_2 \in [0,L-1]}^{p_{i} = \frac{2\pi}{L} n_i }  \frac{4 \sinh \frac{ m }{2} }{ 4  \sinh \frac{E_1}{2}   4\sinh \frac{E_2}{2}} \nonumber \\
 &  \quad\quad  \times  \frac{e^{i ( p_1 x_1 + p_2 x_2 ) }}{ 2 \sinh \frac{E}{2} - ( 2 \sinh \frac{E_1}{2}+ 2 \sinh \frac{E_2}{2})  }, 
\end{align} 
where  $2 \sinh \frac{E_i}{2}=  \sqrt{  2 \cosh m - 2 \cos p_i }$.  In the continuum limit, a relativistic three-body Green's function is obtained
\begin{align}
 &  G_0^D (x_1 , x_2  ; E)  
  \rightarrow  \frac{1}{L^2} \sum_{n_1,n_2 \in \mathbb{Z}}^{p_i=\frac{2\pi}{L} n_i }   \frac{ 2m}{ 2E_1 2E_2}  \frac{e^{i ( p_1 x_1 + p_2 x_2 ) }}{  E- (E_1 + E_2)  } , 
\end{align}
where $E_i = \sqrt{p_i^2+m^2}$.

\paragraph{Coordinate space representation of the lattice version of the quantization condition:} 
 again, by introducing a column vector, $\xi^D(x) = \left [ \Phi^D (x,x) ,\Phi^D (x,0) \right  ]^T$,   a  simple homogeneous equation which is defined in a discrete space is found,
\begin{equation}
\xi^D(x)  =  \sum_{x' \in [0, L-1]}    \mathcal{G}^D(x,x';E)  \xi^D(x'),   \ \  x\in [0,   L-1], 
\end{equation}
where $\mathcal{G}^D$ has a similar expression as its non-relativistic counterpart in  the continuum limit in Eq.~(\ref{Gcornonrel}),  
 \begin{align}
 \mathcal{G}^D_{1,1}(x,x';E)  & =      V_0 G_0^D (x-x', x-x' ; E), \nonumber \\
  \mathcal{G}^D_{1,2}(x,x';E)  & =  2  U_0 G_{(1)}^D (x, x-x' ; E), \nonumber \\
    \mathcal{G}^D_{2,1}(x,x';E)  & =  V_0 G_0^D (-x', x-x' ; E) , \nonumber \\
      \mathcal{G}^D_{2,2}(x,x';E)  & =  U_0  \left [   G_{(1)}^D (  x,- x' ; E)+G_{(2)}^D ( x-x',0 ; E) \right ] .
\end{align}
The  coordinate-space  representation of the quantization condition defined in a discrete finite box  is therefore given by
\begin{align}
& \det \left [ \delta_{\alpha, \beta} \delta_{i,j}   -    \mathcal{G}^D_{\alpha, \beta}(x_i,x_j;\sigma ) \right ] =0, \nonumber \\
& \quad \quad \quad \quad \quad \quad \quad \quad  \ \ (x_i ,x_j) \in [0,\cdots , L-1]. \label{3bquantcorrlat}
\end{align}

\paragraph{Momentum-space representation of the lattice version of the quantization condition:}  in momentum space, using the discrete-space Fourier transformation in Eqs.~(\ref{fourier}-\ref{invfourier}), and also using the identity,
\begin{equation}
\frac{1}{L}\sum_{x \in [0,  L-1]}    e^{- i p x}   = \delta_{p,0}, \ \  p =\frac{2\pi}{L} n, \ n \in [0, L-1], 
\end{equation}
    we find
\begin{equation}
\widetilde{\xi}^D (p)  =   \sum_{n' \in [0,L-1]}^{p' = \frac{2\pi}{L} n'}  \widetilde{\mathcal{G}}^D( p, p';\sigma)  \widetilde{\xi}^D (p')  , 
\end{equation}
where $\ p = \frac{2\pi}{L} n $,  $n \in [0,L-1]$. The kernel function, $\widetilde{\mathcal{G}}^D$, is defined by
\begin{align}
\widetilde{\mathcal{G}}^D_{1,1}( p, p';\sigma) &  =   \delta_{p, p'}    \frac{ V_0 }{L} \sum_{n''  \in  [0,L-1]}^{p'' = \frac{2\pi}{L} n'' }    \widetilde{G}_0^D (p'', p - p'' ; E)      , \nonumber \\
\widetilde{\mathcal{G}}^D_{1,2}( p, p';\sigma) & = ( \frac{ \sinh \frac{E_{p'}}{2}  }{ \sinh \frac{m}{2}} )   \frac{2 U_0}{L}  \widetilde{G}_0^D (p-p', p' ; E)  , \nonumber \\
\widetilde{\mathcal{G}}^D_{2,1}( p, p';\sigma) & =  \frac{ V_0}{L}       \widetilde{G}_0^D(p'-p, p ; E)   , \nonumber \\
\widetilde{\mathcal{G}}^D_{2,2}( p, p';\sigma) &  =   \delta_{p, p'}     (  \frac{ \sinh \frac{E_{p}}{2} }{\sinh \frac{m}{2}} )    \frac{ U_0}{L} \sum_{n''  \in  [0,L-1]}^{p'' = \frac{2\pi}{L} n'' }    \widetilde{G}_0^D (p'', p ; E)    \nonumber \\
&   +  ( \frac{ \sinh \frac{E_{p'}}{2}  }{ \sinh \frac{m}{2}})    \frac{  U_0}{L}    \widetilde{G}_0^D (p, p' ; E)  ,
\end{align}
where  $(p,p') \in \frac{2\pi }{L} n$,   $n \in [0,L-1]$, and 
\begin{align}
 &  \widetilde{G}_0^D (p_1, p_2 ; E)   =   \frac{4 \sinh \frac{ m }{2}  }{ 4  \sinh \frac{E_1}{2}   4\sinh \frac{E_2}{2}} \nonumber \\
 & \times  \frac{1}{ 2 \sinh \frac{E}{2}  - ( 2 \sinh \frac{E_1}{2}+ 2 \sinh \frac{E_2}{2})  } \ .
\end{align}
Hence, the momentum-space representation of the quantization condition in discrete space is given by
\begin{align}
&  \det \left [ \delta_{\alpha, \beta} \delta_{p,p'}   -   \widetilde{ \mathcal{G}}^D_{\alpha, \beta} ( p, p';\sigma)   \right ] =0  ,   \nonumber \\
&\quad \quad \quad \quad   \ \   (p,p') \in \frac{2\pi }{L} n, \   n \in  [0, L-1] \ .  \label{3bquantmomlat}
\end{align}

\begin{figure}
\begin{center}
\includegraphics[width=0.49\textwidth]{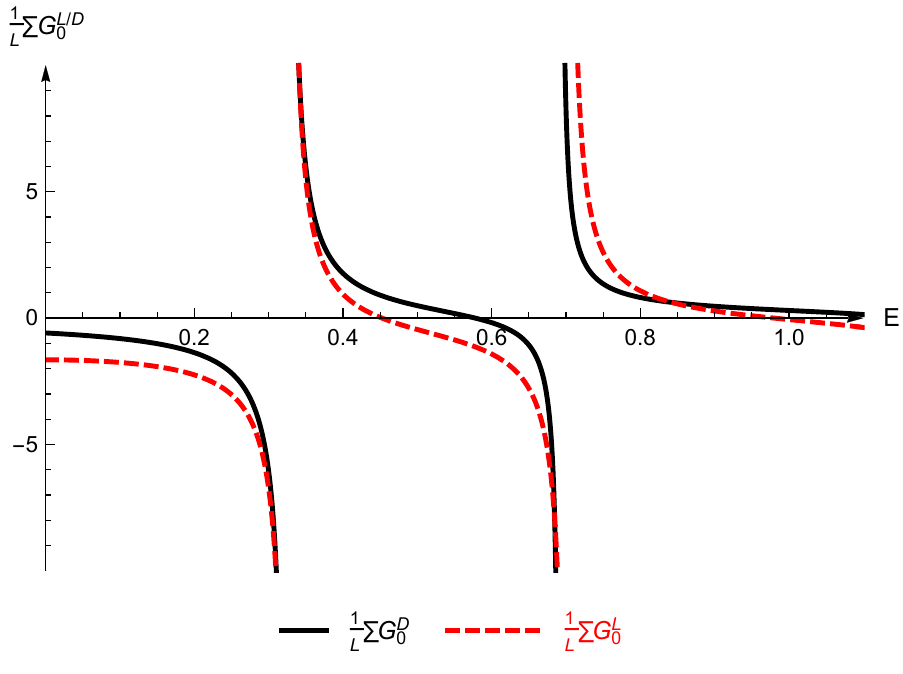}  
\caption{ Plot of the two-body Green's function,  $\frac{ 1 }{L} \sum_{n'  \in  [0,L-1]  }^{p' = \frac{2\pi}{L} n' }    \widetilde{G}_0^D (p', - p' ; E) $  (solid black), as function of total energy $E$   vs. non-relativistic Green's function $\frac{1}{L } \sum_{n \in \mathbb{Z}}^{p' = \frac{2\pi}{L} n'} \widetilde{G}_0^L (p', - p' ; \sigma )   = \frac{\cot \frac{\sigma}{2\sqrt{2}} L }{2 \sqrt{2} \sigma}$ (dashed red), using the lattice energy-momentum dispersion relation, $\cosh \frac{E}{2} = \cosh m + 1 - \cos \frac{\sigma}{\sqrt{2}}$.     The lattice size and mass of the light particle are: $L=20$ and $m=0.163$. \label{G3blatplot}}
\end{center}
\end{figure}

In the non-relativistic and  continuum limits, Eq.~(\ref{3bquantcorrlat}) and Eq.~(\ref{3bquantmomlat}) are reduced to  Eq.~(\ref{3bquantcorr}) and Eq.~(\ref{3bquantmom}).  In the limit    $U_0=0$, in which the interaction between the light  and the heavy static particles vanishes the three-body quantization condition is,  thus, reduced to a simple form,
\begin{equation}
   1= \frac{ V_0 }{L} \sum_{n'  \in  [0,L-1] }^{p' = \frac{2\pi}{L} n' }    \widetilde{G}_0^D (p', p - p' ; E)      , \ \ p = \frac{2\pi}{L} (0,\cdots, L-1).
\end{equation}
The non-relativistic counterpart of the above equation is given by Eq.~(\ref{zeroU0nonrel}), hence $p= \frac{2\pi}{L} (0,\cdots, L-1)$ may be related to the total momentum of the   light-particle system. The comparison of  $\frac{ 1 }{L} \sum_{n'  \in  [0,L-1]  }^{p' = \frac{2\pi}{L} n' }    \widetilde{G}_0^D (p', - p' ; E) $  in the CM frame of the two light particles and its non-relativistic counterpart, $\frac{1}{L } \sum_{n \in \mathbb{Z}}^{p' = \frac{2\pi}{L} n'} \frac{1}{\sigma^2 -{p'}^2}  = \frac{\cot \frac{\sigma}{2\sqrt{2}} L }{2 \sqrt{2} \sigma}$, combined with lattice dispersion relation, $\cosh \frac{E}{2} = \cosh m + 1 - \cos \frac{\sigma}{\sqrt{2}}$, is shown in Fig.~\ref{G3blatplot}.

\begin{figure}
\begin{center}
\includegraphics[width=0.49\textwidth]{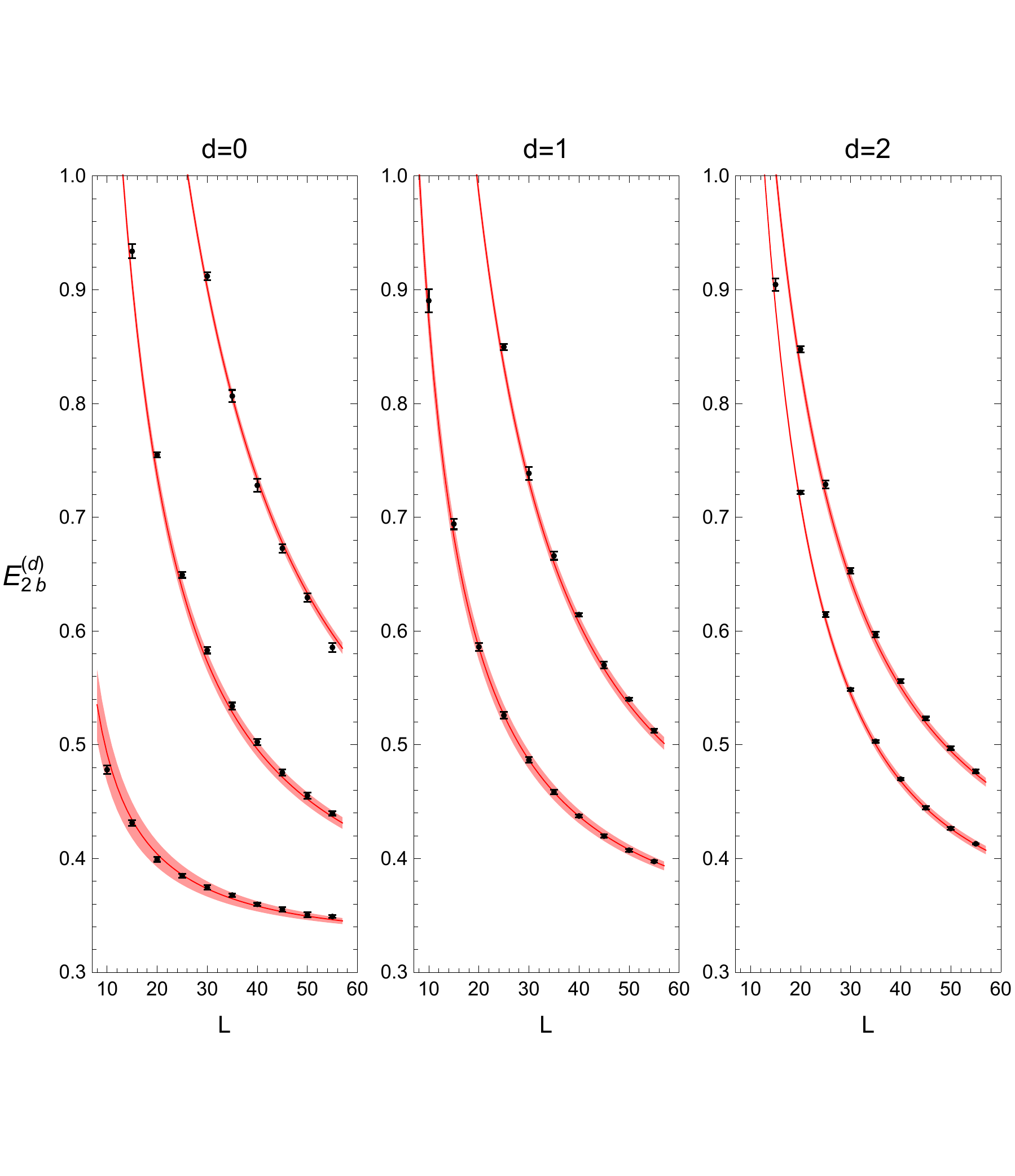}  
\caption{ Plot of  the three-body spectrum from the lattice model simulation (black  data) vs.  fit using the quantization condition in form of Eq.~(\ref{3bquantcorrlat}) or Eq.~(\ref{3bquantmomlat}) (red bands). \label{E2bfitplot}}
\end{center}
\end{figure}

Using the light particle mass  of $m=0.163 \pm 0.001$, and heavy-light short-range potential coupling strength, $U_0 = 0.07 \pm 0.03$  from  the fit to the two-body spectra  as input, the  remaining parameter, i.e., the coupling strength of the short-range potential between  the light particles,  $V_0$, is  determined by fitting  the  two-light-particle spectra   using the three-body quantization condition, either Eq.~(\ref{3bquantcorrlat}) or Eq.~(\ref{3bquantmomlat}). The fitting results are show in Fig.~\ref{E2bfitplot}.  The coupling strength  is determined as  $V_0 = 0.43 \pm 0.03$  through the fit. At last, we would like to remark that due to the   finite lattice spacing $a$, the light particle mass and coupling strengths extracted from the lattice simulation are in principle $a$-dependent as well. The physical value of these quantities in continuum limit may be extrapolated by using  multiple lattice spacing simulation results      combined with the  renormalization group method that yields the explicit lattice spacing dependence of renormalized physical quantities based on perturbation theory.   In principle, the multi-particle dynamics in finite volume and in infinite volume are related though same set of interaction parameters.  Once all the interaction parameters are extracted from lattice results, the physical scattering amplitudes  in infinite volume may be determined and computed though standard procedures, such as Faddeev's approach. The example of  non-relativistic Faddeev's approach is listed in Appendix \ref{faddeevsolinf}.  

\section{Summary }\label{summary}
In summary, a $1+1$ dimensional heavy-light three-body system is simulated by a coupled-channel $\phi^4$-like lattice model. An improved variational approach  for the finite-volume multiparticle  problem is proposed in this work.  The advantage is that the quantization conditions are given in terms of the periodic lattice structure and interaction potentials instead of  explicit finite-volume scattering amplitudes. This opens up the possibility of finding a more practical formalism which can be easily used for data fitting. The interaction potentials are  varied to match the energy spectrum;  once the parameters of the potentials are   extracted by data fitting,   the dynamics of the entire system   is determined. The multiparticle dynamics in infinite volume hence may be computed separately.

\bigskip

\begin{acknowledgements}
 We   acknowledge support from the Department of Physics and Engineering, California State University, Bakersfield, CA.   This research was supported in part by the National Science Foundation under Grant No. NSF PHY-1748958. M.D. acknowledges support by the NSF Career grant No. PHY-1452055. We also thank Akaki Rusetsky for his remarks and fruitful discussions.
 \end{acknowledgements}

\appendix

\section{Fast-converging representation of the periodic Green's function: $G_0^L$} \label{latsumgreen}
The periodic Green's function, $G_0^L$, defined  in Eq.(\ref{G0Leq}), has  the analytic expression
\begin{equation}
 G_0^L (x_1 , x_2  ; \sigma)  = - \frac{i}{4} \sum_{n_1, n_2 \in \mathbb{Z}} H_0^{(1)} (\sigma |  \mathbf{ x} + \mathbf{ n} L |) \ , \label{G0L2}
\end{equation}
where $\mathbf{ x} = (x_1, x_2)$ and $\mathbf{ n} = (n_1, n_2)$.
Equivalently, it can also be expressed as
\begin{equation}
 G_0^L (x_1 , x_2  ; \sigma)  = \frac{1}{L^2} \sum_{n_1, n_2 \in \mathbb{Z}}^{p_{i} = \frac{2\pi}{L} n_i } \frac{e^{i ( p_1 x_1 + p_2 x_2 ) }}{\sigma^2 - p_1^2 - p_2^2} \ . \label{G0L1}
\end{equation}
 Unfortunately, both expressions in Eq.~(\ref{G0L2}) and Eq.~(\ref{G0L1}) suffer   poor convergence, especially for    $\sigma$ values on the real axis.

  \begin{figure}
\begin{center}
\includegraphics[width=0.44\textwidth]{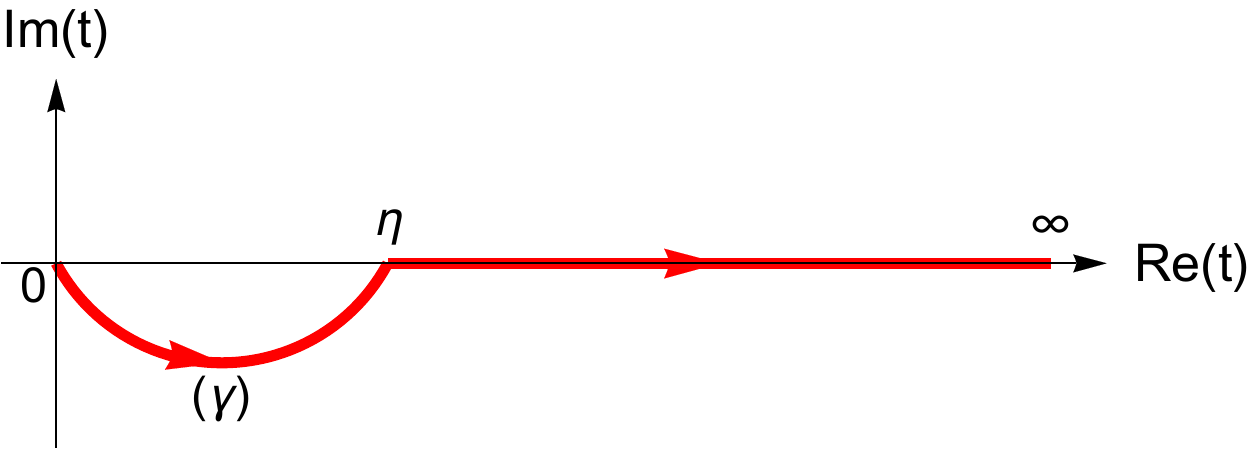}  
\caption{Integration contour of $(\gamma)$ in  the complex $t$ plane. \label{coontourplot}}
\end{center}
\end{figure}

In this section, we present  two equivalent  fast-converging expressions of $G_0^L$.  

(i) First of all  we use the identity 
\begin{equation}
- \frac{i}{4} H_{0}^{(1)} ( \sigma   r ) =  -\frac{1}{2\pi} \left [ \int_{0}^{\eta} (\gamma) + \int_{\eta}^{\infty} \right ] \frac{d t}{t} e^{- r^{2} t^{2} + \frac{\sigma^{2}}{4 t^{2}} }  ,  \label{H0contour}
\end{equation}
where  $\eta$ is an arbitrary parameter. For real  $\sigma$ values, the integration path has to be defined carefully in the complex plane to warrant  the convergence of integration \cite{doi:10.1002/andp.19213690304}, see Fig.~\ref{coontourplot}. Hence, we can rewrite  $G_0^L$ in Eq.~(\ref{G0L2}) as
 \begin{align}
&  G_0^L (x_1 , x_2  ; \sigma)  \nonumber \\
& =    - \frac{i}{4}  H_0^{(1)} (\sigma  | \mathbf{ x} | ) + \frac{1}{2\pi}    \int_{0}^{\eta}  (\gamma) \frac{d t}{t}  e^{- | \mathbf{ x}  |^{2} t^{2} + \frac{\sigma^{2}}{4 t^{2}} }   \nonumber \\
& - \frac{1}{2\pi}    \sum_{ \mathbf{ n }  \in \mathbb{Z}^{2} }^{ \mathbf{ n} \neq \mathbf{ 0} }   \int_{\eta}^{\infty}  \frac{d t}{t}  e^{- | \mathbf{ x} + \mathbf{ n} L |^{2} t^{2} + \frac{\sigma^{2}}{4 t^{2}} }  \nonumber \\
& - \frac{1}{2\pi}  \sum_{ \mathbf{ n }  \in \mathbb{Z}^{2} }     \int_{0}^{\eta}  (\gamma) \frac{d t}{t}  e^{- | \mathbf{ x} + \mathbf{ n} L |^{2} t^{2} + \frac{\sigma^{2}}{4 t^{2}} }  . \label{latsum}
\end{align}
 In this expression, the first three terms are all well-behaved  for  real $\sigma$ values.
For the last term in Eq.~(\ref{latsum}), using the identity
\begin{equation}
\frac{1}{2\pi  } e^{- r^{2} t^{2}} =\frac{1}{2 t^{2}}   \int \frac{d \mathbf{ p}}{(2\pi)^{2}} e^{ - \frac{ \mathbf{p}^{2} }{4 t^{2}}} e^{i \mathbf{ p} \cdot \mathbf{ r}},
\end{equation}
and also applying Poisson summation, we find
 \begin{align}
&    - \frac{1}{2\pi}   \sum_{ \mathbf{ n }  \in \mathbb{Z}^{2} }   \int_{0}^{\eta} d t e^{ -   | \mathbf{ x} +   \mathbf{ n}  L   |^{2} t^{2} + \frac{q^{2}}{4 t^{2}}}  \nonumber \\
&= \frac{1}{L^{2}}   \sum_{ \mathbf{ n} \in \mathbb{Z}^{2}}^{\mathbf{ p} =  \frac{2   \pi}{L} \mathbf{ n}   }  \frac{ e^{ \frac{ \sigma^{2} - \mathbf{ p}^{2}  }{4 \eta^{2}}}  }{\sigma^{2} - \mathbf{ p}^{2} }   e^{i \mathbf{ p}  \cdot \mathbf{ x}  }  .
\end{align}
Putting everything together, we find
  \begin{align}
&  G_0^L (x_1 , x_2  ; \sigma)  =    - \frac{i}{4}  H_0^{(1)} (\sigma  | \mathbf{ x} | )   \nonumber \\
&+ \frac{1}{2\pi}    \int_{0}^{\eta}  (\gamma) \frac{d t}{t}  e^{- | \mathbf{ x}  |^{2} t^{2} + \frac{\sigma^{2}}{4 t^{2}} }  + \frac{1}{L^{2}}   \sum_{ \mathbf{ n} \in \mathbb{Z}^{2}}^{\mathbf{ p} =  \frac{2   \pi}{L} \mathbf{ n}   }  \frac{ e^{ \frac{ \sigma^{2} - \mathbf{ p}^{2}  }{4 \eta^{2}}}  }{\sigma^{2} - \mathbf{ p}^{2} }   e^{i \mathbf{ p}  \cdot \mathbf{ x}  }  \nonumber \\
& - \frac{1}{2\pi}    \sum_{ \mathbf{ n }  \in \mathbb{Z}^{2} }^{ \mathbf{ n} \neq \mathbf{ 0} }   \int_{\eta}^{\infty}  \frac{d t}{t}  e^{- | \mathbf{ x} + \mathbf{ n} L |^{2} t^{2} + \frac{\sigma^{2}}{4 t^{2}} }  . \label{G0Lsum}
\end{align}
The integration in the last term in Eq.~(\ref{G0Lsum}) is normally highly suppressed for $L>10$ and  $\eta > 1$. Therefore, with a modest choice of $\eta \sim 3$, the last piece in Eq.~(\ref{G0Lsum})  can be safely ignored in the numerical evaluation.

(ii) Secondly, using the relation,
\begin{align}
 & G_0^L (x  ; q)   = \frac{1}{L}\sum_{n \in \mathbb{Z}}^{p = \frac{2\pi}{L} n } \frac{e^{i p x}}{q^2 - p^2}  \nonumber \\
& = - \frac{i}{2 \sqrt{q^2}} \left [  e^{i \sqrt{q^2} |x|} + \frac{2 \cos \sqrt{q^2} x}{e^{- i \sqrt{q^2} L}-1} \right ],
\end{align}
the second fast convergent expression of $G_0^L$ can be obtained from Eq.(\ref{G0L1}),
\begin{align}
&   G_0^L (x_1 , x_2  ; \sigma)     \nonumber \\
&  = \frac{1}{L} \sum_{n_1 \in \mathbb{Z}}^{p_{1}= -\frac{2\pi}{L} n_1 }  e^{i  p_1 x_1  }  G_0^L (x_2  ; \sqrt{\sigma^2 - p_1^2})  \nonumber \\
&  = \frac{1}{L} \sum_{n_2 \in \mathbb{Z}}^{p_{2}= -\frac{2\pi}{L} n_2 }  e^{i  p_2 x_2  }  G_0^L (x_1  ; \sqrt{\sigma^2 - p_2^2}) \ .  
\end{align}

\section{Regularization of coordinate space representation  quantization condition} \label{detsub}
Due to the singular nature of  the kernel function, $\mathcal{G}^L_{\alpha, \alpha}(x,x';\sigma)  $   as $x \rightarrow x'$ in Eq.(\ref{3bcoupmat}),  the homogeneous integral equation, Eq.(\ref{3bcoupmat}), defined in continuous space has to be regularized.  In this work, we adopt a modified subtraction quadrature method \cite{kythe2002computational}  for the regularization scheme of singularities. First of all, we can subtract the same term on both side of  Eq.(\ref{3bcoupmat}),
\begin{align}
& \left[ \mathbb{I} - \int_{-\frac{L}{2}}^{\frac{L}{2}} d x' \mathcal{S}(x,x';\sigma) \right ] \xi(x) \nonumber \\
& = \int_{-\frac{L}{2}}^{\frac{L}{2}} d x'  \left [  \mathcal{G}^L(x,x';\sigma)   \xi(x') -     \mathcal{S} (x,x';\sigma)     \xi(x)  \right ],    \label{homosub}
\end{align}
where
\begin{align}
 & \mathcal{S}(x,x';\sigma)  \nonumber  \\
 &  = 
   \begin{bmatrix}     V_0   G_{0}^L(x-x', x-x';\sigma)  &  0 \\ 
0&   U_0   G_{0}^L(0, x-x';\sigma) 
   \end{bmatrix}. 
\end{align}
Using relations,
  \begin{align}
  &  \int_{- \frac{L}{2}}^{ \frac{L}{2}}    d  x'  G_0^L(x- x'; x- x'; \sigma)       =\frac{\cot \frac{  \sigma  L}{2\sqrt{2}} }{2 \sqrt{2}  \sigma } ,   \\
   &    \int_{- \frac{L}{2}}^{ \frac{L}{2}}    d  x'  G_0^L( 0 ; x- x'; \sigma)        = \frac{\cot \frac{  \sigma  L }{2}}{2   \sigma } , 
 \end{align}
 we can discretize  Eq.(\ref{homosub}) to
\begin{align}
& \begin{bmatrix}    1-   V_0 \frac{\cot \frac{  \sigma  L}{2\sqrt{2}} }{2 \sqrt{2}  \sigma }   &  0 \\ 
0&  1-   U_0    \frac{\cot \frac{  \sigma  L }{2}}{2   \sigma } 
   \end{bmatrix}  \xi(x_i) \nonumber \\
& = \sum_{j} w_j  \left [  \mathcal{G}^L(x_i ,x_j;\sigma)   \xi(x_j) -     \mathcal{S} (x_i,x_j;\sigma)     \xi(x_i)  \right ].     \label{homoreg}
\end{align}
Hence, all the singular terms cancel out as $x_j = x_i$ on the left hand side above equation, discretized Eq.(\ref{homoreg}) now is well-defined. Reorganizing Eq.(\ref{homoreg}), we find
\begin{equation}
\sum_j   \mathcal{D}(x_i, x_j; \sigma)   \xi(x_j)  =0, \label{detreg}
\end{equation}
where
\begin{align}
 & \mathcal{D}_{1,1}(x_i, x_j; \sigma) \nonumber \\
 &  = \delta_{i,j} \left [ 1-  V_0 \frac{\cot \frac{  \sigma  L}{2\sqrt{2}} }{2 \sqrt{2}  \sigma } + \sum_{k \neq i } w_k  \mathcal{S}_{1,1}(x_i,x_k;\sigma)    \right ] \nonumber \\
 & - w_j    V_0 G_{0}^L(x_i-x_j, x_i -x_j;\sigma)|_{i \neq j} ,
\end{align}
\begin{align}
 & \mathcal{D}_{2,2}(x_i, x_j; \sigma) \nonumber \\
 &  = \delta_{i,j} \left [ 1-   U_0    \frac{\cot \frac{  \sigma  L }{2}}{2   \sigma }  + \sum_{k \neq i } w_k  \mathcal{S}_{2,2}(x_i,x_k;\sigma)    \right ] \nonumber \\
 & - w_j     U_0 \left [  G_{0}^L(0,x_i-x_j;\sigma)|_{i \neq j} + G_0^L(-x_j, x_i ; \sigma) \right ] ,
\end{align}
and
\begin{align}
 \mathcal{D}_{\alpha,\beta}(x_i, x_j; \sigma)  = - w_j   \mathcal{G}^L_{\alpha,\beta}(x_i,x_j;\sigma) , \ \ \mbox{if} \ \  \alpha \neq \beta.
\end{align}
The non-trivial solution of homogenous equation, Eq.(\ref{detreg}), exist, provided  
\begin{equation}
\det \left [   \mathcal{D} (x_i, x_j; \sigma) \right ] =0.
\end{equation}

\section{Reduction of Bethe-Salpeter equation to relativistic Schr\"odinger equation}\label{reductionBS}
\subsection{ Reduction of two-body system}
Let's consider the general form of Bethe-Salpeter equation \cite{Salpeter:1951sz} for one heavy and one light scalar particles bound state system,
\begin{equation}
 \psi_{BS} (q) = \frac{-i}{(p_1^2 - m^2)  (p_2^2 - M^2) }  \int \frac{ d^4 k}{(2\pi)^4} I (k-q) \psi_{BS} (k ) ,
\end{equation}
where $p_1= (p_{1,0}, \mathbf{ p}_1)$ and $p_2= (p_{2,0}, \mathbf{ p}_2)$ are the four momenta of two particles,  $m$ and $M$ represent mass of light and heavy particles respectively. $q=\frac{M p_1 - m p_2}{m+M}$ denotes the relative four momentum of two particles. $\psi (q)$ is the Bethe-Salpeter wave function, and  $I(k)$ represent ladder approximation interaction kernel function.  Using approximation of "instantaneous interaction  kernel", $I(k) = I(\mathbf{ k})$, in which interaction kernel has no dependence on the energy components of four momenta, and also introducing the Schr\"odinger equation wave function by $\psi(\mathbf{ q}) = \int  \frac{ d q_0 }{2\pi} \psi_{BS}(q)$, we obtain,
 \begin{equation}
 \psi (\mathbf{ q}) =   \int  \frac{ d q_0 }{2\pi}  \frac{-i}{(p_1^2 - m^2)  (p_2^2 - M^2) }  \int \frac{ d \mathbf{ k}}{(2\pi)^3} I (\mathbf{ k}-\mathbf{ q}) \psi (\mathbf{ k}) .
\end{equation}
Next, using identity
\begin{equation}
 \int \frac{d q_0}{2\pi}  \frac{-i}{(p_1^2 - m^2)  (p_2^2 - M^2) }   =   \frac{1}{2E_1 2 E_2} \frac{2 (E_1 +E_2)}{E^2- (E_1+E_2)^2},
\end{equation}
where $E_1 = \sqrt{  \mathbf{ p}_1^2+m^2}$,  $E_2 = \sqrt{  \mathbf{ p}_2^2+M^2}$ and $E$ is total energy of two-particle system. In the heavy static limit, $E_2 \rightarrow M \rightarrow \infty$ and $\mathbf{ p}_1 \rightarrow \mathbf{ q}$,  we also shift total energy by $M$ to  subtract the heavy mass, $E \rightarrow E -M$, and introducing potential by $\widetilde{V} = \frac{1}{2 M} I $,  then we find
 \begin{equation}
 \psi (\mathbf{ q}) = \frac{1}{ 2 \sqrt{\mathbf{ q}^2 + m^2}  } \frac{1}{E-  \sqrt{\mathbf{ q}^2 + m^2} }    \int \frac{ d \mathbf{ k}}{(2\pi)^3} \widetilde{V} (\mathbf{ k}-\mathbf{ q}) \psi (\mathbf{ k}) . \label{rel2bBS}
\end{equation}
In non-relativistic limit, Eq.(\ref{rel2bBS}) hence yield familiar Lippmann-Schwinger equation, 
 \begin{equation}
 \psi (\mathbf{ q}) =   \frac{1}{\sigma^2-   \mathbf{ q}^2  }    \int \frac{ d \mathbf{ k}}{(2\pi)^3} \widetilde{V} (\mathbf{ k}-\mathbf{ q}) \psi (\mathbf{ k}) . 
\end{equation}

\subsection{ Reduction of three-body system}
Now, let's consider  one heavy and two light scalar particles bound state system, the Bethe-Salpeter equation is given by
\begin{align}
&  \psi_{BS} (q_1, q_2) = \frac{(-i)^2}{(p_1^2 - m^2)  (p_2^2 - m^2)  (p_3^2 - M^2) }  \nonumber \\
 & \times \int \frac{ d^4 k_1}{(2\pi)^4}   \frac{ d^4 k_2}{(2\pi)^4} I (k_1-q_1, k_2 - q_2) \psi_{BS} (k_1 , k_2 ) ,
\end{align}
where $p_i = (p_{i,0}, \mathbf{ p}_i)$   are the four momenta of three particles,  particle-1 and -2 are labeled as light particles and particle-3 denotes heavy particle. The relative momenta are introduced by  $q_1=\frac{ (M+m) p_1 - m (p_2+p_3 )}{2 m+M}$  and $q_2=\frac{ (M+m) p_2 - m (p_1+p_3 )}{2 m+M}$. Again, assuming  "instantaneous interaction  kernel", $I (k_1, k_2 )   = I(\mathbf{ k}_1, \mathbf{ k}_2 )$, and introducing Schr\"odinger equation wave function, $\psi  (\mathbf{ q}_1, \mathbf{ q}_2)  = \int \frac{d q_{1,0}}{2\pi}  \frac{d q_{2,0}}{2\pi} \psi_{BS} (q_1, q_2) $, hence, we get
\begin{align}
&  \psi  (\mathbf{ q}_1, \mathbf{ q}_2)  =  \int \frac{d q_{1,0}}{2\pi}  \frac{d q_{2,0}}{2\pi}  \frac{(-i)^2}{(p_1^2 - m^2)  (p_2^2 - m^2)  (p_3^2 - M^2) }  \nonumber \\
 & \times \int \frac{ d  \mathbf{ k}_1}{(2\pi)^3}   \frac{ d \mathbf{ k}_2}{(2\pi)^3}  I(\mathbf{ k}_1-\mathbf{ q}_1, \mathbf{ k}_2-\mathbf{ q}_2 \psi  (\mathbf{ k}_1, \mathbf{ k}_2) .
\end{align}
The integration over propagators can be carried out by
\begin{align}
&  \int \frac{d q_{1,0}}{2\pi}  \frac{d q_{2,0}}{2\pi}  \frac{(-i)^2}{(p_1^2 - m^2)  (p_2^2 - m^2)  (p_3^2 - M^2) } \nonumber \\
&  =   \frac{1}{2E_1 2 E_2 2 E_3} \frac{2 (E_1 +E_2 + E_3)}{E^2- (E_1+E_2 +E_3)^2},
\end{align}
where  $E_{1,2} = \sqrt{  \mathbf{ p}_{1,2}^2+m^2}$ and  $E_3 = \sqrt{  \mathbf{ p}_3^2+M^2}$.

When only pair-wise interactions are considered, the three-body interaction potential, $  I$, may be given by sum of interactions among all pairs,
\begin{align}
& I(\mathbf{ k}_1-\mathbf{ q}_1, \mathbf{ k}_2-\mathbf{ q}_2 ) \nonumber \\
& = \sum_{(i,j) =1,2}^{j \neq i} 2 E_j (2\pi)^3 \delta(\mathbf{ k}_j -\mathbf{ q}_j  )  I_U(\mathbf{ k}_i -\mathbf{ q}_i  )  \nonumber \\
 &+ 2 E_3 (2\pi)^3 \delta(\mathbf{ k}_1 +  \mathbf{ k}_2 -\mathbf{ q}_1-\mathbf{ q}_2   )  I_V ( \frac{\mathbf{ k}_1-\mathbf{ k}_2 }{2} - \frac{ \mathbf{ q}_1-\mathbf{ q}_2 }{2}).
\end{align}
The relativistic kinematic factors $\langle \mathbf{ p}_i | \mathbf{ p}'_i \rangle=2 E_{i} (2\pi)^3 \delta(\mathbf{ p}_i -\mathbf{ p}'_i  )$ emerge   when i-th particle  is free propagating and  not involved in the  interaction.
At the limit of static heavy particle, $M \rightarrow \infty$, $\mathbf{ p}_i \rightarrow \mathbf{ q}_i$,  and   define potentials by $\widetilde{U} = \frac{1}{  2 M} I_U $ and $\widetilde{V} = \frac{1}{2m } I_V $, we find
\begin{align}
&  \psi  (\mathbf{ q}_1, \mathbf{ q}_2)  =   \frac{1}{2  E_1 2  E_2 } \frac{1}{E- (E_1+E_2 )}  \nonumber \\
 & \times \bigg [ (2 E_2 ) \int \frac{ d  \mathbf{ k}_1}{(2\pi)^3}     \widetilde{U}(\mathbf{ k}_1-\mathbf{ q}_1  )  \psi  (\mathbf{ k}_1, \mathbf{ q}_2) \nonumber \\
 &  \quad    + (2 E_1 ) \int \frac{ d  \mathbf{ k}_2}{(2\pi)^3}     \widetilde{U}(\mathbf{ k}_2-\mathbf{ q}_2  )  \psi  (\mathbf{ q}_1, \mathbf{ k}_2) \nonumber \\
 & \quad +   (2 m ) \int \frac{ d  \mathbf{ k}}{(2\pi)^3}     \widetilde{V}(\mathbf{ k}-  \mathbf{ q}_1  )  \psi  (  \mathbf{ k}  , \mathbf{ q}_1+\mathbf{ q}_2  -\mathbf{ k}  ) \bigg ] . \label{rel3bBS}
\end{align}
If all interactions are turned off except interaction between pair (13), above equation is thus reduced to
\begin{align}
  \psi  (\mathbf{ q}_1, \mathbf{ q}_2)&  =   \frac{1}{2  E_1  } \frac{1}{(E-E_2) - E_1 }   \nonumber \\
 & \times  \int \frac{ d  \mathbf{ k}_1}{(2\pi)^3}     \widetilde{U}(\mathbf{ k}_1-\mathbf{ q}_1  )  \psi  (\mathbf{ k}_1, \mathbf{ q}_2)  ,
\end{align}
which is just a  relativistic two-body LS equation in Eq.(\ref{rel2bBS}) with second particle as a spectator. 

In non-relativistic limit, Eq.(\ref{rel3bBS})   yield a familiar form,
\begin{align}
&  \psi  (\mathbf{ q}_1, \mathbf{ q}_2)  =  \frac{1}{ \sigma^2 - \mathbf{ q}_1^2  - \mathbf{ q}_2^2}  \nonumber \\
 & \times  \bigg [   \int \frac{ d  \mathbf{ k}_1}{(2\pi)^3}     \widetilde{U}(\mathbf{ k}_1-\mathbf{ q}_1  )  \psi  (\mathbf{ k}_1, \mathbf{ q}_2) \nonumber \\
 &  \quad    +  \int \frac{ d  \mathbf{ k}_2}{(2\pi)^3}     \widetilde{U}(\mathbf{ k}_2-\mathbf{ q}_2  )  \psi  (\mathbf{ q}_1, \mathbf{ k}_2) \nonumber \\
 & \quad +    \int \frac{ d  \mathbf{ k}}{(2\pi)^3}     \widetilde{V}(\mathbf{ k}-  \mathbf{ q}_1  )  \psi  (  \mathbf{ k}  , \mathbf{ q}_1+\mathbf{ q}_2  -\mathbf{ k}  ) \bigg ] .  
\end{align}

\section{Solutions of heavy-light three-body system in infinite space}\label{faddeevsolinf}
In infinite space, the dynamics of a non-relativistic heavy-light three-body system is described by Schr\"odinger equation, 
\begin{equation}
\left [ \sigma^2 + \hat{T} - U_0 \delta(x_1) - U_0 \delta (x_2) - V_0\delta(r) \right ] \Psi(x_1,x_2)=0, 
\end{equation}
where  $\sigma^2 = 2 m E$, $ \hat{T} = \nabla_{x_1}^2+ \nabla_{x_2}^2$,  and $r=x_1 - x_2$. For scattering solutions, the wave function has the following form \cite{Faddeev:1960su,9780706505740}, $\Psi = \Psi_{(0)} + \sum_{\gamma=1}^3 \Psi_{(\gamma)}$, where $\Psi_{(0)}$ stands for the incoming free wave, and  $\Psi_{(\gamma)}$ satisfy coupled equations,
\begin{align}
& \Psi_{(i)}(x_1,x_2) = \int d x'_1 d x'_2 G_{(i)} (x_1,x_2; x'_1,x'_2; \sigma)  \nonumber \\
& \times U_0 \delta(x'_i) \left [ \Psi_{(0)}  (x'_1,x'_2) + \sum_{\gamma \neq i} \Psi_{(\gamma)}  (x'_1,x'_2)\right ], \ \ i=1,2, \label{lipmannlight}
\end{align}
and
\begin{align}
& \Psi_{(3)}(x_1,x_2) = \int d x'_1 d x'_2 G_{(3)} (x_1,x_2; x'_1,x'_2; \sigma)  \nonumber \\
& \times V_0 \delta(x'_1 - x'_2) \left [ \Psi_{(0)}  (x'_1,x'_2) + \sum_{\gamma  = 1}^2 \Psi_{(\gamma)}  (x'_1,x'_2)\right ],   \label{lipmannheavy}
\end{align}
where the Green's functions are the solutions of equations,
\begin{align}
& \left [ \sigma^2 + \hat{T} - U_0 \delta(x_i) \right ]  G_{(i)} (x_1,x_2; x'_1,x'_2; \sigma) \nonumber \\
&  \quad \quad \quad \quad  = \delta(x_1-x'_1) \delta(x_2 - x'_2),  \ \ i = 1,2,
\end{align}
and
\begin{align}
& \left [ \sigma^2 + \hat{T} - V_0 \delta(x_1-x_2) \right ]  G_{(3)} (x_1,x_2; x'_1,x'_2; \sigma) \nonumber \\
&  \quad \quad \quad \quad  = \delta(x_1-x'_1) \delta(x_2 - x'_2) .
\end{align}
The analytic expressions of Green's functions are
\begin{align}
& G_{(1)} (x_1,x_2; x'_1,x'_2; \sigma) = - \int \frac{d p}{2\pi} \frac{i e^{i p (x_2-x_2')}}{2 \sqrt{\sigma^2 - p^2}} \nonumber \\
& \times \left [ e^{ i   \sqrt{\sigma^2 - p^2}  |x_1-x'_1|} + i t_{U} ( \sqrt{\sigma^2 - p^2} ) e^{i  \sqrt{\sigma^2 - p^2}  (|x_1|+|x'_1|)} \right ], \label{G1faddeev}
\end{align}
where $t_{U} (k) = - \frac{U_0}{2  k + i U_0}$ is heavy-light two-body scattering amplitude, the expression of $G_{(2)}$ is obtained by exchanging the role of $x_1 \leftrightarrow x_2$ in Eq.(\ref{G1faddeev}), and
\begin{align}
& G_{(3)} (x_1,x_2; x'_1,x'_2; \sigma) = - \int \frac{d p}{2\pi} \frac{i e^{i p (R-R')}}{4 \sqrt{\frac{\sigma^2}{2} - \frac{p^2}{4}}} \nonumber \\
& \times \left [ e^{ i  \sqrt{\frac{\sigma^2}{2} - \frac{p^2}{4}} |r-r'|} + i t_{V} (\sqrt{\frac{\sigma^2}{2} - \frac{p^2}{4}}) e^{i \sqrt{\frac{\sigma^2}{2} - \frac{p^2}{4}} (|r|+|r'|)} \right ], \label{G3faddeev}
\end{align}
where $R = \frac{x_1+x_2}{2}$ and $r=x_1-x_2$ are  CM   and relative position of two light particles, and $t_{V} (k) = - \frac{V_0}{4  k  + i V_0} $ stands for the light-light two-body scattering amplitude.

By introducing scattering $T$-amplitude,
\begin{align}
T_{(1)} (p) &= -  U_0 \int  d x   e^{- i p x} \Psi(0, x),  \nonumber \\
T_{(2)} (p) &= -  U_0 \int  d x   e^{- i p x} \Psi( x,0),  \nonumber \\
T_{V} (p) & = -  V_0 \int  d x   e^{- i p x} \Psi(x, x), 
\end{align}
and also using Lippmann-Schwinger equations, Eq.(\ref{lipmannlight}-\ref{lipmannheavy}), and explicit expression of Green's function in Eq.(\ref{G1faddeev}-\ref{G3faddeev}), we find
\begin{align}
\Psi_{(1)} (x_1,x_2)  & = \int \frac{d p}{2\pi} \frac{i e^{i  \sqrt{\sigma^2 - p^2} |x_1|} e^{i p x_2}}{2 \sqrt{\sigma^2 - p^2}}   T_{(1)} (p) , \nonumber \\
\Psi_{(2)} (x_1,x_2)  & = \int \frac{d p}{2\pi} \frac{i e^{i  \sqrt{\sigma^2 - p^2} |x_2|} e^{i p x_1}}{2 \sqrt{\sigma^2 - p^2}}   T_{(2)} (p) ,
\end{align}
and
\begin{equation}
\Psi_{(3)} (x_1,x_2)  =  \int \frac{d p}{2\pi} \frac{i e^{ i  \sqrt{\frac{\sigma^2}{2} - \frac{p^2}{4}} |r|}  e^{i p R }}{4 \sqrt{\frac{\sigma^2}{2} - \frac{p^2}{4}}} T_{V} (p).
\end{equation}
The $T$ amplitudes satisfy coupled integral equations,
\begin{align}
& \frac{ T_{(i)} (p) -  T^{(0)}_{(i)} (p) }{ 2 \sqrt{\sigma^2 -p^2} t_U ( \sqrt{\sigma^2 -p^2} )}  \nonumber \\
&=  -  \int \frac{d k}{2\pi} \left [  \frac{T_{(j)} (k)}{\sigma^2 - p^2 -k^2}  +  \frac{T_{V} (k)}{\sigma^2 - \frac{k^2}{2} -2 (p-\frac{k}{2})^2} \right ] , 
\end{align}
where $(i,j) = 1,2$  $(i\neq j)$, and
\begin{align}
 \frac{ T_{V} (p) -  T^{(0)}_{V} (p) }{ 4   \sqrt{\frac{\sigma^2}{2} - \frac{p^2}{4}}  t_V (  \sqrt{\frac{\sigma^2}{2} - \frac{p^2}{4}}  )}  =  -  \int \frac{d k}{2\pi}  \frac{T_{(1)} (k) + T_{(2)} (k)}{\sigma^2 - k^2 - (k-p)^2}      ,
\end{align}
where
\begin{align}
\frac{    T^{(0)}_{(1)} (p) }{ 2 \sqrt{\sigma^2 -p^2} t_U ( \sqrt{\sigma^2 -p^2} )}  & = \int d x e^{- i p x} \Psi_{(0)} (0,x), \nonumber \\
\frac{    T^{(0)}_{(2)} (p) }{ 2 \sqrt{\sigma^2 -p^2} t_U ( \sqrt{\sigma^2 -p^2} )}  & = \int d x e^{- i p x} \Psi_{(0)} (x,0), \nonumber \\
 \frac{  T^{(0)}_{V} (p) }{ 4   \sqrt{\frac{\sigma^2}{2} - \frac{p^2}{4}}  t_V (  \sqrt{\frac{\sigma^2}{2} - \frac{p^2}{4}}  )}  & =    \int d x e^{- i p x}  \Psi_{(0)} (x,x)  .
\end{align}

  \begin{figure}
\begin{center}
\includegraphics[width=0.48\textwidth]{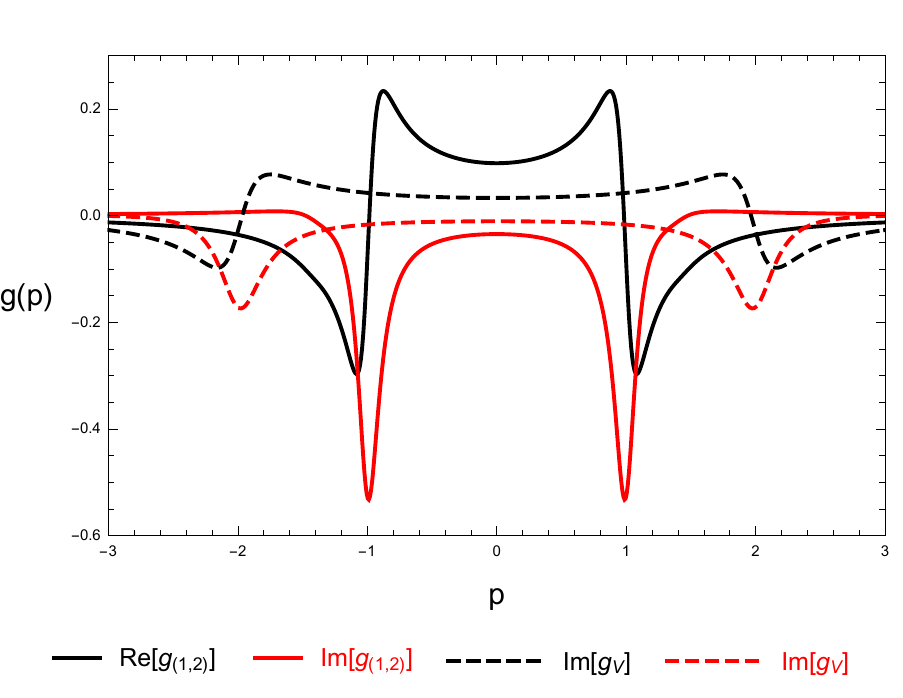}  
\caption{Numerical solutions of $g$ amplitudes in Eq.(\ref{gUeq}-\ref{gVeq}), with $p_1=-p_2 = 1+  0.1 i   $,  a small imaginary part is given to $p_1$ to smooth out the curves near the pole position for a better visualization purpose. Black   and red colors are assigned to represent real part and imaginary part of $g$ amplitudes respectively, and solid and dashed curves are associated to $g_{(1,2)}$ and $g_V$ amplitudes respectively. \label{gampplot}}
\end{center}
\end{figure}

\subsection{Scattering solutions of Faddeev equation with an incoming plane  wave: $\Psi_{(0)} (x_1,x_2) = e^{i p_1 x_1} e^{i p_2 x_2}$}
For the scattering solutions of Faddeev equation with an an incoming plane  wave: $\Psi_{(0)} (x_1,x_2) = e^{i p_1 x_1} e^{i p_2 x_2}$, where $p_i$ stands for the incoming momentum of $i$-th light particle, we have
\begin{align}
   T^{(0)}_{(i)} (p)   & = 2 \sqrt{ p_i^2 } t_U (\sqrt{ p_i^2 }) (2\pi) \delta(p-p_j), \ \ (i \neq j) = 1,2, \nonumber \\
 T^{(0)}_{V} (p)   & = 4  \sqrt{q^2}   t_V ( \sqrt{q^2}   ) ( 2\pi ) \delta(p - P),
\end{align}
where $q= \frac{p_1 - p_2}{2}$ and $P= p_1+ p_2$. Usually, singular terms, $T^{(0)}$'s, are not convenient for numerical computation, so it is better to introduce $g$-amplitude by a shift, 
\begin{align}
& g_{(i)} (p) =   \frac{ T_{(i)} (p) -  T^{(0)}_{(i)} (p)   }{ 2 \sqrt{\sigma^2 -p^2} t_U ( \sqrt{\sigma^2 -p^2} )} -  Q_{(i)} (p) , \nonumber \\
&g_V (p) =  \frac{ T_{V} (p) -  T^{(0)}_{V} (p) }{ 4   \sqrt{\frac{\sigma^2}{2} - \frac{p^2}{4}}  t_V (  \sqrt{\frac{\sigma^2}{2} - \frac{p^2}{4}}  )}  ,
\end{align}
where $Q_{(i)}$'s satisfy  coupled equations, 
\begin{align}
& Q_{(i)} (p)  = - \frac{2 \sqrt{p_j^2} t_U (\sqrt{p_j^2} )}{p_j^2 -p^2}  \nonumber \\
& - \int \frac{d k}{2\pi} \frac{2 \sqrt{\sigma^2 - k^2} t_U (\sqrt{\sigma^2 -k^2})}{\sigma^2 - p^2 -k^2} Q_{(j)} (k) ,  \ \ (i \neq j) = 1,2.
\end{align}
The analytic expressions of $Q_{(i)}$'s are given by
\begin{align}
& 2 \sqrt{\sigma^2 -p^2} t_U ( \sqrt{\sigma^2 -p^2} ) Q_{(i)} (p)  \nonumber \\
&  =-  \frac{ 2 \sqrt{p_1^2} t_U (\sqrt{p_1^2} )  2 \sqrt{p_2^2} t_U (\sqrt{p_2^2} ) }{p_j^2 -p^2} , \ \ (i \neq j ) = 1,2.
\end{align}

The $g$ amplitudes are the solutions of coupled equations that are free of $\delta$-function type singularities,
\begin{align}
 & g_{(i)} (p)  =   g^{(0)}_{(i)} (p)  \nonumber \\
 & -  \int \frac{d k}{2\pi} \frac{2 \sqrt{\sigma^2 - k^2} t_U (\sqrt{\sigma^2 -k^2})}{\sigma^2 - p^2 -k^2} g_{(j)} (k)  \nonumber \\
 &  -  \int \frac{d k}{2\pi} \frac{4 \sqrt{ \frac{ \sigma^2 }{2}- \frac{k^2}{4}} t_V (\sqrt{ \frac{ \sigma^2 }{2} -\frac{k^2}{4}})}{\sigma^2 - \frac{k^2}{2} - 2 (p - \frac{k}{2} )^2} g_{V} (k), \ \ (i \neq j ) = 1,2, \label{gUeq}
\end{align}
and
\begin{align}
 & g_{V} (p)  =   g^{(0)}_{V} (p)   \nonumber \\
 & -  \int \frac{d k}{2\pi} \frac{2 \sqrt{\sigma^2 - k^2} t_U (\sqrt{\sigma^2 -k^2})}{\sigma^2 - k^2 - (k-p)^2}  \left [  g_{(1)} (k)  +  g_{(2)} (k)  \right ] , \label{gVeq}
\end{align}
where
\begin{align}
& g^{(0)}_{(1,2)} (p)   = - \frac{ 2 \sqrt{q^2} t_V (\sqrt{q^2})}{ q^2 - (p- \frac{P}{2})^2},  \nonumber \\
& g^{(0)}_{V} (p)  = - \sum_{i=1,2}^{ j\neq i} \frac{2 \sqrt{p_i^2} t_U ( \sqrt{p_i^2}  ) }{p_i^2 - (p-p_j)^2}  \nonumber \\
& + i  t_U ( \sqrt{p_1^2}  )  t_U ( \sqrt{p_2^2}  ) \sum_{i=1,2}^{ j\neq i}   \frac{2  \sqrt{p_j^2} }{p^2 - 4 p_i^2} . 
\end{align}

In terms of $g$ amplitudes, the wave function of three-body heavy-light system is thus given by
\begin{align}
& \Psi(x_1,x_2) = \psi_{p_1} (x_1) \psi_{p_2} (x_2) + i t_V(\sqrt{q^2}) e^{i \sqrt{q^2} |r| } e^{i P R} \nonumber \\
& + \sum_{i=1,2}^{j\neq i} \int \frac{d p}{2\pi} e^{ i \sqrt{\sigma^2 - p^2} | x_i |} e^{i p x_j } i t_U (\sqrt{\sigma^2 - p^2}  ) g_{(i)} (p) \nonumber \\
& + \int \frac{d p}{2\pi} e^{ i \sqrt{ \frac{ \sigma^2 }{2}- \frac{p^2}{4}} | r |} e^{i p R } i t_V ( \sqrt{ \frac{ \sigma^2 }{2}- \frac{p^2}{4}}  ) g_{V} (p), \label{waveinf}
\end{align}
where $ \psi_{p} (x)  = e^{i p x} + i t_U (\sqrt{p^2}) e^{i \sqrt{p} |x|}$ is heavy-light two-body wave function. The first term on the right-hand side of Eq.(\ref{waveinf}), $\psi_{p_1} (x_1) \psi_{p_2} (x_2) $, stands for the solution of heavy-light quark system at the limit of zero  interaction between two light particles, $V_0=0$. The second term represent the disconnected contribution of two light particles interacting while heavy particle play the role of spectator. The last two terms in Eq.(\ref{waveinf}) produce diffraction effect  that are generated by the rescattering effect among all the pairs.

Integral equations of $g$ amplitudes can be solved with standard matrix inversion method, see  Fig.\ref{gampplot} for   examples of numerical solutions of Eq.(\ref{gUeq}-\ref{gVeq}). 

\bibliography{ALL-REF.bib}

\end{document}